\documentclass{article} 
\PassOptionsToPackage{table}{xcolor} 
\usepackage{iclr2027_conference,times}

\usepackage{amsmath,amsfonts,bm}

\def\eqref#1{equation~\ref{#1}}

\def\1{\bm{1}}

\DeclareMathAlphabet{\mathsfit}{\encodingdefault}{\sfdefault}{m}{sl}
\SetMathAlphabet{\mathsfit}{bold}{\encodingdefault}{\sfdefault}{bx}{n}

\usepackage[utf8]{inputenc} 
\usepackage[T1]{fontenc}    
\usepackage{hyperref}       
\usepackage{url}            
\usepackage{booktabs}       
\usepackage{amsfonts}       
\usepackage{nicefrac}       
\usepackage{microtype}      

\usepackage{subcaption}
\usepackage[utf8]{inputenc} 
\usepackage[T1]{fontenc}    
\usepackage{hyperref}       
\usepackage{enumitem}
\usepackage{url}            
\usepackage{booktabs}       
\usepackage{graphicx}
\usepackage{booktabs}
\usepackage{amsmath}
\usepackage{amsfonts}       
\usepackage{nicefrac}       
\usepackage{microtype}      
\usepackage{amssymb}
\usepackage{multirow}
\usepackage{adjustbox}
\usepackage{rotating}
\usepackage{wrapfig}
\usepackage{booktabs,multirow,adjustbox}

\usepackage{tcolorbox}
\tcbuselibrary{breakable, skins}
\newtcolorbox{promptbox}[1][]{
    breakable,
    enhanced,
    colback=gray!8,
    colframe=gray!50,
    fonttitle=\bfseries\small,
    title=#1,
    left=6pt, right=6pt, top=4pt, bottom=4pt,
    arc=3pt
}

\usepackage{pifont}
\newcommand{\cmark}{\textcolor{green!70!black}{\ding{51}}}%
\newcommand{\xmark}{\textcolor{red}{\ding{55}}}%

\newcommand{\header}[1]{\vspace{-3pt}\noindent{\bf #1}}

\usepackage[textsize=tiny,textwidth=3.3cm]{todonotes}

\newcommand{\textblue}[1]{\textcolor{blue}{#1}}

\newif\ifshowdiff\showdifftrue
\definecolor{NewText}{HTML}{0B5FA5}   
\definecolor{NewRow}{HTML}{E8F1FA}    

\newcommand{\name}{{OmniDream}}
\newcommand{\studynumber}{20}

\title{Enabling Immersive Audio-Visual Experience\\ from Any Video}

\author{Zitong Lan, Mutian Tong, Jiatao Gu, Mingmin Zhao \\
University of Pennsylvania\\
}
\iclrfinalcopy 
\begin{document}

\maketitle

\begin{abstract}

Most videos capture only a narrow field of view and provide  no spatial audio, limiting the sense of immersion they can provide.
Recent video generation models can expand perspective videos into panoramic ones, but do not provide the corresponding spatial soundscape.
Without spatially consistent audio, these expanded visual worlds remain incomplete.
This paper presents \name{}, a training-free framework that transforms a silent monocular video into an immersive audiovisual experience, where viewers can freely look around while sounds remain spatially aligned with the visual scene.
At the core of \name{} is an object-centric audio representation that disentangles each sound source's intrinsic audio content from its scene-dependent acoustic effects, enabling independent audio generation, physics-based simulation of propagation effects, and flexible spatial audio rendering.
Experiments show improved audio-visual alignment, spatial correctness, and perceptual immersiveness over baselines.
Examples are available on \href{https://huggingface.co/spaces/CuriousAlien000/spatial-audio-360-demo}{\textblue{Hugging Face}}.
\end{abstract}

\vspace{-10pt}
\begin{figure}[h]
    \centering
    \vspace{-6pt}
    \includegraphics[width=0.85\linewidth]{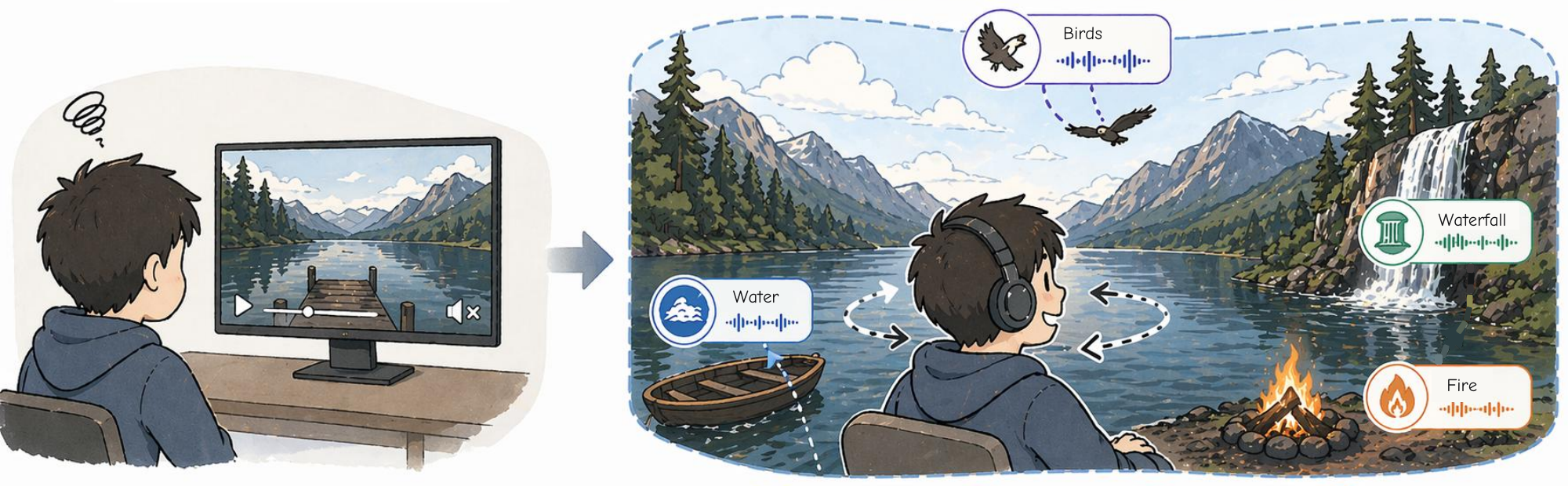}
    \vspace{-10pt}
    \caption{\name{} transforms a silent perspective video into a $360^\circ$ video with synchronized spatial audio. As users turn their head and inspect different parts of the scene, both the visual perspective and perceived sound change consistently with the environment.}
    \vspace{-15pt}
    \label{fig:teaser}
\end{figure}

\section{Introduction}
\label{sec:introduction}
\vspace{-5pt}

Most videos today offer a passive viewing experience. Viewers are restricted to the recorded viewing direction, while their soundtracks are often absent or non-spatial.
As a result, users remain observers of a scene rather than participants within it.
In contrast, immersive media such as $360^\circ$ video with spatial audio allows users to turn their heads and explore different parts of the environment while sounds remain naturally and spatially anchored to the scene.
For instance, as the user turns toward a waterfall, the rushing water grows louder and the pitch gets higher; as a train passes by, its sound moves across the scene, with directional cues reflecting its changing position.
Together, these visual and auditory cues create a substantially stronger sense of presence and realism than ordinary videos.

Despite growing support for immersive playback (e.g., $360^\circ$ video and spatial audio on YouTube), immersive content remains scarce~\citep{imvid2025, tan2024imagine360}.
This is because capturing rich scene information typically requires specialized sensing equipment, including panoramic cameras, microphone arrays, and RF sensors, along with carefully calibrated acquisition pipelines~\cite{morgado2018spatialaudio,liu2025omniaudio,surfradar,lai2026non}.
In contrast, ordinary perspective videos are abundant on the internet and increasingly easy to create with video generation models~\citep{wan2025,kong2024hunyuanvideo}.
This raises a question: \emph{can we transform an ordinary video into an immersive audio-visual experience?}


Existing methods address separate components of immersive audio-visual generation.
Video-to-audio models synthesize sound for perspective videos~\citep{chen2025video, cheng2025mmaudio, liu2024tell, luo2023diff, tian2025audiox, wang2025audiogen}, while spatial audio methods additionally provide directional cues~\citep{karchkhadze2025stereofoley, kim2025visage, liu2025omniaudio, xie2026sonic4d, zhang2025visaudio, zhao2025foleyspace}; neither expands the visual scene beyond the input's field of view.
Conversely, perspective-to-$360^\circ$ video generation expands the visual environment but leaves the audio unaddressed~\citep{tan2024imagine360, fang2025viewpoint, luo2025beyond, li2026cubecomposer}.

Bringing these capabilities together requires an audio representation that connects individual sound sources to the expanded visual scene.
This is challenging because an immersive soundscape must respond to changes in the listener's orientation while keeping sound sources anchored to the scene.
A pre-rendered binaural mixture cannot adapt on its own to arbitrary listener orientations, while first-order ambisonics (FOA) supports rotation but limits directional resolution through a first-order spherical harmonic approximation~\citep{benhur2021binaural,frank2015producing}.
Additionally, generating directly into these formats ties the audio representation to the chosen playback format.

We address this challenge through \textit{object-centric audio generation and rendering}: we generate a dry audio track for each sounding object, simulate its source-to-listener impulse response (IR), and render its acoustic contribution before combining the sources into the final soundscape. This IR captures propagation delays, attenuation, and reverberation determined by source location, scene geometry, and materials~\citep{luo2022learning,wang2024hearing,lan2024acoustic}.
This formulation offers multiple advantages: (i) \textbf{object-level audio-visual grounding}, generating each dry track from its corresponding visual object and associating it with an explicit source trajectory in the 3D scene reconstructed from video, accounting for object motion and changes in camera viewpoint; (ii)  \textbf{physics-grounded acoustic propagation modeling}, computing each source's IR through acoustic simulation in this reconstructed 3D scene using the source trajectories and estimated material properties; and (iii) \textbf{steerable and flexible spatial rendering}, retaining individual sources for rendering at different listening orientations and into different output formats, including binaural audio and ambisonics~\citep{frank2015producing}.

Building on this formulation, we present \name{}, a training-free framework that transforms a silent perspective video into a panoramic video with spatial audio.
\name{} improves overall spatial correlation between audio and visual content by about 52\% relative to the strongest baselines.
In a study with 20 participants, \name{} improves spatial correctness ratings by about 34\% and overall immersiveness ratings by about 28\% relative to the strongest baseline.
Ablations show that object-centric tracking and full acoustic rendering contribute to spatial correspondence. Against measured IRs, the zero-shot renderer has a multi-resolution STFT error of 0.58.

In summary, our contributions are as follows:
\begin{itemize}[leftmargin=10pt, itemindent=0pt, itemsep=2pt, topsep=-2pt, parsep=0pt]
\item We present the first framework for immersive audio-visual generation that transforms a silent perspective video into a $360^\circ$ video with spatially and temporally aligned audio.
\item We introduce a novel object-centric formulation of audio generation and rendering that combines sound synthesis with physically simulated propagation, enabling scene-grounded, steerable audio.
\item We extensively evaluate our system through objective metrics and a user study, demonstrating improvements over baseline methods. Robustness analyses and ablation studies further assess its reliability and the contribution of each component.
\end{itemize}

\vspace{-10pt}
\section{Related Work}
\vspace{-15pt}

\suppressfloats[t]
Tab.~\ref{tab:comparison} compares existing methods in terms of their visual and audio outputs, audiovisual synchronization, physics-aware rendering, and support for separate object audio tracks.

\begin{table}[t]
\centering
\resizebox{.99\linewidth}{!}{
\setlength{\tabcolsep}{6pt}
\begin{tabular}{lccccccc}
\toprule
\textbf{Method} &
\textbf{Input}&
\textbf{Visual Output}&
\textbf{Audio Output}&
\textbf{\shortstack{Audio-Visual\\Synchronization}} &
\textbf{\shortstack{Physics-Aware\\Rendering} }&
\textbf{\shortstack{Multi-Track\\Audio}}\\
\midrule
MMAudio~\citep{cheng2025mmaudio} & Video & Video & Mono & \cmark & \xmark & \xmark \\
Sonic4D~\citep{xie2026sonic4d} & Video & Video &  Stereo & \cmark & \xmark & \xmark \\
StereoFoley~\citep{karchkhadze2025stereofoley}  & Video & Video & Stereo & \cmark &\xmark & \xmark \\
ViSAGe~\citep{kim2025visage} & Video & Video & FOA & \cmark & \xmark & \xmark \\
OmniAudio~\citep{liu2025omniaudio} & $360^\circ$ video & $360^\circ$ video & FOA & \cmark & \xmark & \xmark \\
See2Sound~\citep{dagli2025see} & Image & Image & 5.1 Sound & \xmark & \xmark & \cmark \\
AVObject~\citep{li2025sounding} & Image & Image & Mono & \xmark & \xmark & \cmark \\
SonoWorld~\citep{jin2026sonoworld} & Image & 3D static scene & FOA & \xmark &\xmark & \cmark \\
Argus~\citep{luo2025beyond} & Video & $360^\circ$ video  &  No Audio & \xmark & \xmark & \xmark \\
CubeComposer~\citep{li2026cubecomposer} & Video & $360^\circ$ video  &  No Audio & \xmark & \xmark & \xmark \\
\textbf{\name{}} & Video  & $360^\circ$ video & FOA & \cmark & \cmark & \cmark \\
\bottomrule
\end{tabular}}
\vspace{-5pt}
\caption{
    Comparison of immersive media generation methods.
    First-Order Ambisonics (FOA) is a spatial audio format with direction-dependent sound perception.
    \textit{Physics-Aware Rendering} indicates whether the method models geometry-based acoustic propagation and environmental effects.
    \textit{Multi-Track Audio} indicates whether the method decomposes the scene into individual sounding objects.
    }
\label{tab:comparison}
\vspace{-12pt}
\end{table}

\header{Video Outpainting and Generation.}
Video outpainting aims to extend the spatial extent of a video beyond its original field of view while maintaining temporal consistency and visual coherence. Early work adapted image outpainting techniques to the video domain through propagation-based or generative inpainting strategies \citep{10.1145/2980179.2982398, 9857385}, and more recent approaches leverage diffusion-based video generation models to produce high-quality extrapolations with improved temporal stability\citep{chen2025infinite, 11094145, zheng2026next}.
A related line of research tackles the more ambitious goal of generating full $360^\circ$ panoramic video from a narrow perspective input~\citep{tan2024imagine360, fang2025viewpoint, luo2025beyond, li2026cubecomposer,wu360anything}.
These methods demonstrate impressive visual results, yet the generated contents carry no information about audio .
In our work, the expanded panoramic video serves not only as the final visual output but also as the geometric and semantic basis for immersive audio generation.

\header{Video-to-Audio Generation.}
Video-to-audio generation studies how to synthesize plausible soundtracks from visual input.
Recent methods have shown a strong ability to generate semantically meaningful and temporally aligned audio from silent videos~\citep{chen2025video, cheng2025mmaudio, liu2024tell, luo2023diff, tian2025audiox, wang2025audiogen}.
However, most of these approaches focus on non-spatial audio and operate only within the original camera view.
As a result, they do not model where sounds originate in space or how the surrounding environment shapes what is heard.
In addition, existing works also explore generating audio from images~\citep{jin2026sonoworld, dagli2025see, li2025sounding}.
However, these models lack video-audio synchronization, and the visual experience remains static.
In contrast, our goal is to lift a silent video into an immersive experience with temporally and spatially aligned visual and audio contents.

\header{Spatial Audio Generation.}
Spatial audio methods generate directional sound in binaural~\citep{zhang2025visaudio,zhao2025foleyspace, lan2025guiding, lan2026smartdj}, stereo~\citep{sun2024both}, or FOA formats~\citep{heydari2025immersediffusion}.
These methods condition on perspective video~\citep{gao20192, li2024cyclic,zhang2025visaudio,kim2025visage, karchkhadze2025stereofoley, xie2026sonic4d, zhao2025foleyspace, zhang2025isdrama} or panoramic video, as in OmniAudio~\citep{liu2025omniaudio}.
Pre-rendered binaural audio cannot adapt on its own to head rotation, while FOA supports rotation with limited directional resolution due to its first-order spherical harmonic approximation~\citep{benhur2021binaural,frank2015producing}. Our formulation retains individual sources and their acoustic responses for rendering into different output formats.
Existing methods rely on the supplied visual scene; we expand a silent perspective video into a panoramic scene and simulate sound propagation through its reconstructed geometry.

\vspace{-10pt}
\section{Method}
\label{sec:method}
\vspace{-15pt}

\begin{figure}[t]
    \centering
    \includegraphics[width=.85\linewidth]{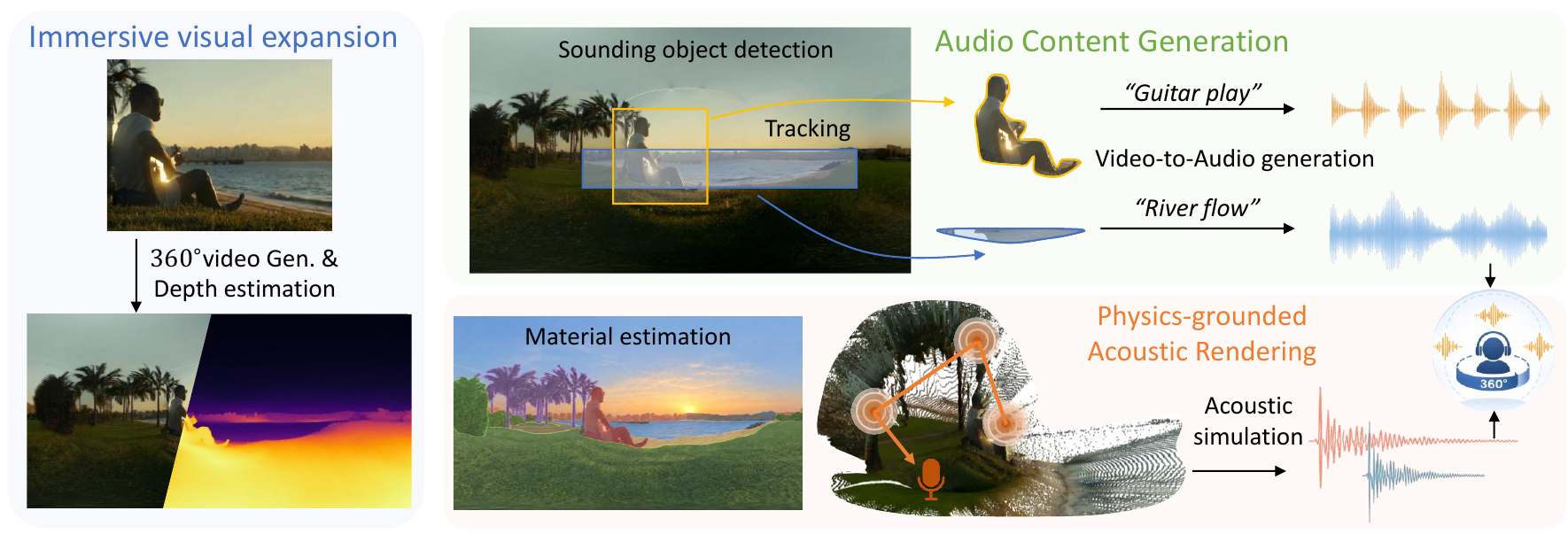}
    \vspace{-10pt}
    \caption{\textbf{\name{} pipeline}. We first expand the narrow perspective video to $360^\circ$ video and reconstruct the scene geometry (Sec.~\ref{sec:visual}). Based on this, we analyze the sounding object in the scene and track their positions through the video and generate per-object dry sound (Sec.~\ref{sec:audio}). Finally, we estimate the acoustic properties on the extracted scene mesh and perform ray tracing to simulate the acoustic effect for each object independently and render the final spatial audio (Sec.~\ref{sec:acoustic}).
    }
    \vspace{-15pt}
    \label{fig:pipeline}
\end{figure}

Given a silent monocular perspective video, \name{} generates a temporally consistent panoramic video $\mathbf{P}$ together with first-order ambisonic (FOA) audio $\mathbf{A}$ that matches with the video.
Fig.~\ref{fig:pipeline} illustrates the full pipeline, which comprises three components.
\textbf{Immersive Visual Expansion} (\S\ref{sec:visual}) expands the narrow field of view into a full panoramic scene and reconstructs the surrounding 3D geometry.
\textbf{Audio Content Generation} (\S\ref{sec:audio}) identifies individual sounding objects in the expanded scene and generates a separate dry audio track for each of them.
\textbf{Physics-Grounded Acoustic Rendering} (\S\ref{sec:acoustic}) renders each dry sound into final wet sound using impulse responses simulated from the recovered scene geometry.
This decomposition grounds spatialization in scene structure and enables interpretable, controllable immersive synthesis.

\vspace{-6pt}
\subsection{Immersive Visual Expansion}
\label{sec:visual}
\vspace{-10pt}
This component expands the current perspective view and constructs a geometrically consistent panoramic scene.
The resulting scene serves three roles: the visual output delivered to the viewer, the visual context for per-object audio generation, and the geometric foundation for acoustic simulation.

%
\header{Panoramic video expansion.}
We use an off-the-shelf model~\citep{li2026cubecomposer} to generate a panoramic video $\mathbf{P}\!=\!\{p_t\}_{t=1}^{T}$ from $\mathbf{V}$.
The downstream audio pipeline operates on the panoramic video and recovered scene representation rather than a specific visual model, making \name{} compatible with different visual frontends that provide these outputs.



\header{Depth estimation for mesh reconstruction.}
To recover the 3D layout of the surrounding environment, we apply a panoramic video depth estimation model~\citep{lin2025depth} to each frame of $\mathbf{P}$, producing a per-frame depth map $\{d_t\}_{t=1}^T$.
In our pipeline, we treat these per-frame predictions as relative depth and use a Vision Language Model (VLM)~\citep{comanici2025gemini} to estimate a metric scale for the scene.
To extract the mesh from the depth images, we back-project $d_t$ into a point cloud $\mathcal{C}_t$ for each frame $t$, and triangulate it by connecting neighboring pixels in the equirectangular grid.
To avoid bridging foreground and background across depth discontinuities, we discard any triangle whose vertex depths differ by more than a threshold $\tau$, yielding an open triangle mesh $\mathcal{M}_t$ that represents only visible geometry.
In addition, since the scene evolves through time, we construct a separate mesh per frame, resulting in a sequence of geometry $\{\mathcal{M}_t\}_{t=1}^T$ for accurate simulations.
Please refer to App.~\ref{sec:app_qual} for more qualitative results on depth estimation.

\vspace{-5pt}
\subsection{Audio Content Generation}
\label{sec:audio}
\vspace{-10pt}
Given the object-centric audio representation, this component generates the audio content of each sounding object.
We identify and track sounding regions in the panoramic video, then synthesize a separate dry audio track conditioned on each object's visual events and semantic description.

\begin{wrapfigure}[15]{r}{0.38\linewidth}
    \centering
    \vspace{-10pt}
    \includegraphics[width=\linewidth]{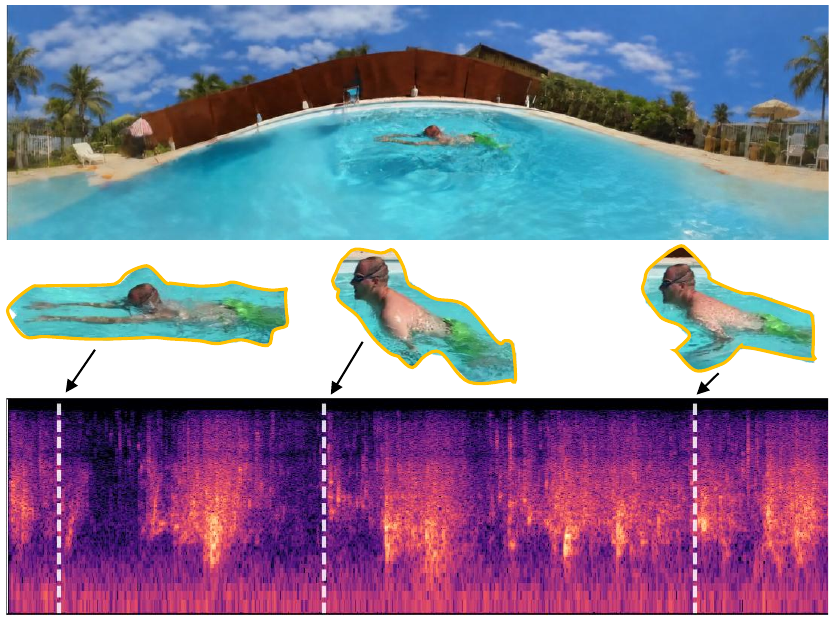}
    \vspace{-15pt}
    \caption{Object tube keyframes and the STFT spectrogram of the generated audio for a swimmer, showing accurate audio-visual timing alignment.}
    \vspace{-10pt}
    \label{fig:video_to_audio}
\end{wrapfigure}
\header{Sounding object proposal.}
We use a VLM~\citep{comanici2025gemini} to analyze the panoramic video $\mathbf{P}$ and propose candidate sounding objects.
We sample representative keyframes and prompt the VLM to identify objects that are likely to produce sound, describing each with a semantic audio description $c_k$ (e.g.\ "waterfall falling", "dog barking", "volcano is erupting lava to the air", "a monster roaring") and a spatial bounding box $b_k$ in the panoramic image plane.
Formally, the VLM produces a set of $K$ proposals $\mathcal{B} = \{(b_k, c_k)\}_{k=1}^K$, where $c_k$ is the audio description of the $k$-th sounding object.

\header{Sounding object tracking and segmentation.}
Given the bounding box proposals in $\mathcal{B}$, we use a video segmentation model (SAM2~\citep{ravi2024sam}) to propagate each proposal into a dense spatiotemporal segmentation mask.
SAM2 takes each bounding box $b_k$ on a reference frame and tracks the corresponding object region through the full video, yielding a sequence of masks $\{m_k^t\}_{t=1}^T$ for each object $k$.
To better capture the interaction between a sounding object and its surroundings (e.g.\ water splashing against rocks, or footsteps on a surface), we expand each mask before extracting the object tube.
We apply morphological dilation with a disk-shaped structuring element of radius $r$ to each binary mask $m_k^t$, yielding an expanded mask $\tilde{m}_k^t = m_k^t \oplus \mathcal{D}_r$, where $\oplus$ denotes the dilation operator and $\mathcal{D}_r$ is a disk of radius $r$.
We extract an object tube $\mathbf{O}_k = \{\pi_k^t(p_t \odot \tilde{m}_k^t)\}_{t=1}^T$, so that each tube captures the object together with its interactions with the environment.

\header{Per-object audio synthesis.}
Each object tube $\mathbf{O}_k$ and its corresponding semantic audio description $c_k$ is passed to a video-to-audio generation model \citep{cheng2025mmaudio} to produce a monaural audio $s_k$.
By conditioning on isolated object tubes rather than the full panoramic frame, audio generation is guided separately to each source, reducing interference from other visual content (Fig.~\ref{fig:video_to_audio}).

\header{Reverberation screening.}
Rendering assumes each source is dry sound, i.e., carrying no or little acoustic effect, since convolving wet sound with an IR would create extra acoustic effect.
MMAudio output is monaural and unpanned but not guaranteed anechoic, so we screen each $s_k$ with the pretrained blind RT60 estimator released with Visual Acoustic Matching~\citep{chen2022visual} and regenerate any source exceeding a reverberation time of 0.15\,s. We regenerate rather than dereverberate, as dereverberation would inject artifacts into a signal that is subsequently convolved; Sec.~\ref{sec:ablation} quantifies the residual reverberation.

\vspace{-10pt}
\subsection{Physics-Grounded Acoustic Rendering}
\label{sec:acoustic}
\vspace{-10pt}

With per-object audio signals and the reconstructed scene geometry in hand, the final step is to render each source into spatial audio by simulating how sound propagates through the physical environment.

\header{Acoustic property assignment.}
In addition to geometry, acoustic simulation requires material properties.
To assign these properties, we first apply SAM2 to each frame $p_t$ of $\mathbf{P}$ in semantic segmentation mode, which segments the scene into $S$ material regions $\{\mathcal{R}_s^t\}_{s=1}^S$.
\begin{wrapfigure}[11]{r}{0.44\linewidth}
    \centering
    \vspace{-12pt}
    \includegraphics[width=\linewidth]{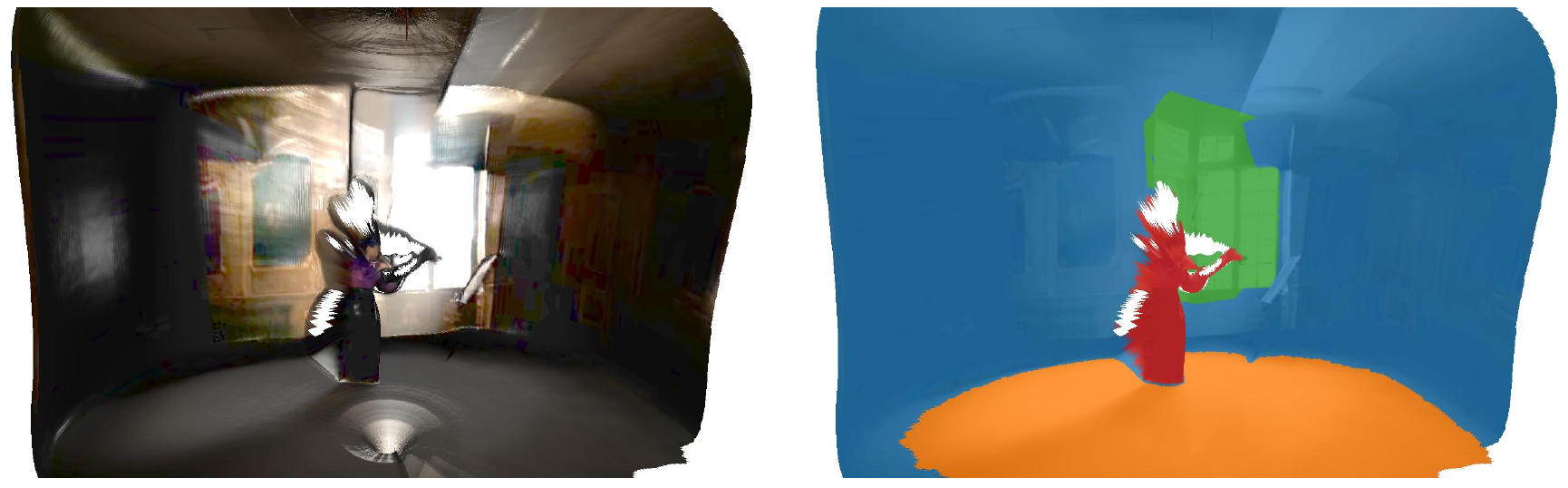}
    \vspace{-15pt}
    \caption{An example of scene geometry and material segmentation. Left is the raw mesh and right is material-annotated mesh, where colors indicate surface regions with different acoustic properties assigned by the VLM.}
    \vspace{-10pt}
    \label{fig:material_seg}
\end{wrapfigure}
We then prompt a proprietary VLM~\citep{comanici2025gemini} again with these segmented regions and query it for acoustic properties including reflection coefficient $\alpha_s$ and scattering coefficient $\sigma_s$ for each material class $s$ at six octave bands.
Since the material categories of static surfaces are stable across frames, the VLM is queried once per unique material label and the resulting $(\alpha_s, \sigma_s)$ pairs are cached and re-used.
We assign these parameters to the corresponding faces of each per-frame mesh $\mathcal{M}_t$, producing the material-annotated geometry used for acoustic ray tracing (Fig.~\ref{fig:material_seg}).
See Sec.~\ref{sec:app_impl} for additional details.

\header{Acoustic effect simulation.}
For each frame $t$, we lift the centroid of each tracked sounding object using the depth map $d_t$ to estimate its 3D source position $\mathbf{x}_k^t$.
AcoustiX~\citep{lan2024acoustic, lan2026resounding, lan2026building, zheng2025scalable} uses these positions and the material-annotated mesh to simulate a per-frame impulse response for each object.
Keeping source content separate from spatial encoding allows the same tracks to be rendered in different formats, including higher-order ambisonics, without regenerating the audio.
We use FOA for compatibility with immersive playback platforms and players, and for direct comparison with FOA baselines.
FOA format encodes a sound field using four channels, including an omnidirectional channel $W$ and three channels $X$, $Y$, $Z$ aligned to the Cartesian axes.
For a source at position $\mathbf{x}_k^t$ relative to the listener, the FOA encoding weights are given by the zeroth-order and first-order spherical harmonics evaluated at the source direction $\hat{\mathbf{x}}_k^t = \mathbf{x}_k^t / \|\mathbf{x}_k^t\|$:
$
    \boldsymbol{\gamma}(\hat{\mathbf{x}}_k^t) = \bigl[1,\;\!\hat{x}_k^t,\;\!\hat{y}_k^t,\;\!\hat{z}_k^t\bigr]^\top.
$
AcoustiX traces a set of rays from $\mathbf{x}_k^t$ through $\mathcal{M}_t$, accumulating specular reflections weighted by $(1\!-\!\alpha_s)$ and diffuse scattering weighted by $\sigma_s$ at each surface interaction.
Each arriving wavefront at the listener is weighted by $\boldsymbol{\gamma}$ according to its angle of incidence, producing a per-frame four-channel FOA impulse response $\mathbf{h}^t_k\!\in\!\mathbb{R}^{4\times L}$:
\begingroup
\setlength{\abovedisplayskip}{4pt}
\setlength{\belowdisplayskip}{4pt}
\begin{equation}
    \mathbf{h}_k^t[\ell] = \sum_{r} g_{k,r}^t\, \boldsymbol{\gamma}(\hat{\mathbf{d}}_{k,r}^t)\, \delta[\ell - \ell_{k,r}^t],
    \label{eq:point_ir}
\end{equation}
\endgroup
where $\ell$ indexes discrete delay samples, the sum is over all simulated ray paths $r$, $g_{k,r}^t$ is the path gain incorporating geometric and propagation absorption at frame $t$, $\hat{\mathbf{d}}_{k,r}^t$ is the direction of arrival, and $\ell_{k,r}^t$ is the arrival delay of path $r$ in samples.
The simulation runs for each sounding object, with the listener placed at the panoramic camera position.

\header{Spatially extended sources.}
Our default renderer approximates each sounding object as a point source located at the centroid of its region. Sources such as crowds, waterfalls, and ocean waves instead occupy an extended region and should not collapse to a single direction.
Let $\Omega_k^t$ denote the 3D source region obtained by lifting the tracked object mask with depth. We sample $M_k$ emitter locations $\{\mathbf{x}_{k,m}^t\}_{m=1}^{M_k}$ uniformly with respect to surface area over $\Omega_k^t$ and simulate a per-emitter FOA IR $\mathbf{h}_{k,m}^t$ using Eq.~\ref{eq:point_ir}. These responses form an effective regional IR:
\begingroup
\setlength{\abovedisplayskip}{4pt}
\setlength{\belowdisplayskip}{4pt}
\begin{equation}
    \bar{\mathbf{h}}_k^t[\ell]
    = \sum_{m=1}^{M_k} w_{k,m}
      \bigl(\mathbf{h}_{k,m}^t * q_{k,m}\bigr)[\ell],
    \qquad
    \sum_{m=1}^{M_k} w_{k,m}^2 = 1,
    \label{eq:regional_ir}
\end{equation}
\endgroup
where $q_{k,m}$ is an all-pass filter that preserves the magnitude spectrum while altering the phase, ensuring that signals from the sampled emitters do not combine coherently.
The weight $w_{k,m}$ controls each emitter's contribution; for uniform sampling, we use $w_{k,m}=1/\sqrt{M_k}$ to preserve the source energy. This superposition distributes energy across multiple directions to create a diffuse sound field. In the following, $\tilde{\mathbf{h}}_k^t$ denotes $\mathbf{h}_k^t$ for a point source and $\bar{\mathbf{h}}_k^t$ for a spatially extended source.

\header{Spatial audio rendering and mixing.}
Rendering and mixing operate on the impulse responses of the chosen output format.
Here, each source signal $s_k\!\in\!\mathbb{R}^{1 \times N}$ is rendered using its effective FOA impulse response $\tilde{\mathbf{h}}_k^t$.
Since convolution is linear, the per-emitter responses in Eq.~\ref{eq:regional_ir} can be aggregated before applying the source signal. Both point and regional sources are therefore rendered similarly:
\begingroup
\setlength{\abovedisplayskip}{4pt}
\setlength{\belowdisplayskip}{4pt}
\begin{equation}
    \mathbf{a}_k^t[n] = \sum_{\ell=0}^{L-1} \tilde{\mathbf{h}}_k^t[\ell]\, s_k[n - \ell]
    \;,
    \label{eq:foa_rendering}
\end{equation}
\endgroup
where $n$ indexes discrete time samples within the frame. Each channel of $\mathbf{a}_k^t$ carries the corresponding FOA component ($W$, $X$, $Y$, $Z$), encoding source direction and the reverberation induced by the instantaneous scene geometry $\mathcal{M}_t$.
The final FOA mixture at frame $t$ is obtained by summing across all $K$ sources: $\mathbf{A}^t[n]\!=\!\sum_{k=1}^K \mathbf{a}_k^t[n]$.
The full FOA track $\mathbf{A}\!=\!\{\mathbf{A}^t\}_{t=1}^T$ is assembled by concatenating across frames with cross-fade effect to ensure smooth audio transition between frames.
This yields an immersive four-channel audio track in which each sounding object contributes energy from its physically correct direction.
The final output is $(\mathbf{P},\!\mathbf{A})$: a temporally consistent $360^\circ$ panoramic video and a synchronized and physics-grounded FOA audio track.

\vspace{-10pt}
\section{Experiments}
\label{sec:experiments}
\vspace{-10pt}
\subsection{Experimental Setup}
\label{sec:setup}
\vspace{-5pt}

\header{Evaluation Dataset.}
We evaluate on two test sets:
\begin{itemize}[leftmargin=10pt, itemindent=0pt, itemsep=2pt, topsep=-2pt, parsep=0pt]
    \item \textbf{ImmerseSet}: We curate 65 in-the-wild monocular perspective videos, spanning indoor environments, outdoor natural settings, and urban scenes, each 8 seconds long without audio.
    50 clips come from the Pexels stock-video library and 15 from current text-to-video models; we exclude clips with severe camera motion, jitter, or blur, which degrade outpainting and depth estimation upstream.
    Since no ground-truth panoramic video or reference audio is available, we assess semantic, temporal and spatial alignment between the generated audio and visual contents.
    \item \textbf{YT360}: A dataset of $360^\circ$ videos with FOA audio~\citep{morgadoNIPS20}; we use a held-out subset of 1000 videos.
    It has reference recordings so we can measure audio similarity metrics. Since $360^\circ$ video is provided, this dataset directly assesses video-to-FOA audio generation.
\end{itemize}

\header{Baselines.}
Since no existing method performs the full perspective-to-immersive task, each baseline addresses a subset of it:
\textbf{MMAudio}~\citep{cheng2025mmaudio}, a video-to-audio model producing mono audio that we replicate to four channels as pseudo-FOA;
\textbf{MMAudio (Spatial)}, a stronger directional variant we construct by running MMAudio on four directional views of the panorama and encoding each track into FOA at its view direction;
\textbf{See2Sound}~\citep{dagli2025see}, image-to-spatial-audio;
\textbf{ViSAGe}~\citep{kim2025visage}, perspective-video-to-FOA;
and \textbf{OmniAudio}~\citep{liu2025omniaudio}, panoramic-video-to-FOA.
App.~\ref{sec:app_baselines} details the input each baseline receives.

\header{Evaluation Metrics.}
Spatial metrics (CC, AUC) are adopted from ViSAGe~\citep{kim2025visage} and measure the spatial and temporal correspondence between a generated audio energy map and a reference spatial map.
We report CC/AUC over the full clip and at 1\,fps and 5\,fps; the harder 5\,fps setting tests whether the audio tracks spatial change rather than merely placing energy in the right region.
For audio-visual alignment, the ImageBind score measures semantic similarity in a shared embedding space, while DeSync estimates the temporal offset between the generated audio and video using Synchformer~\citep{iashin2024synchformer}.
Audio similarity metrics (KL, FD) are standard reference-based measures of audio distribution, computed against ground-truth FOA recordings and therefore applied only on the YT360 evaluation set. Please find more details in App.~\ref{sec:app_metrics}.

\vspace{-5pt}
\subsection{Main Results}
\label{sec:results}
\label{sec:user_study}
\vspace{-5pt}

\begin{figure}[t]
    \centering
    \begin{minipage}[t]{0.6\linewidth}
    \centering
    \vspace{-15pt}
    \captionsetup{hypcap=false}
    \includegraphics[width=\linewidth]{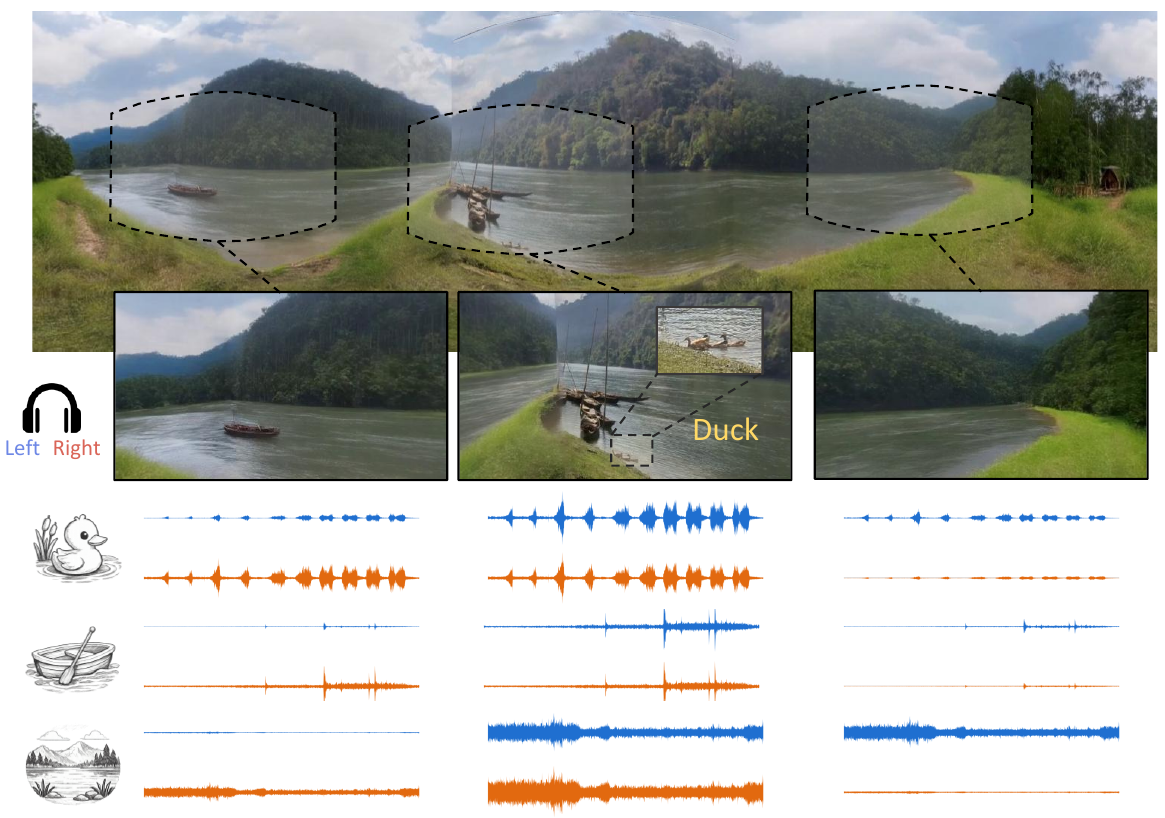}
    \vspace{-15pt}
    \captionof{figure}{Example of generated visual scene with audios. When users look at different perspective views, the spatial audio will also change for different sounding object tracks.}
    \label{fig:example_visualization}
    \end{minipage}
    \hspace{0.1\linewidth}%
    \begin{minipage}[t]{0.215\linewidth}
    \centering
    \vspace{-15pt}
    \captionsetup{hypcap=false}
    \includegraphics[width=\linewidth]{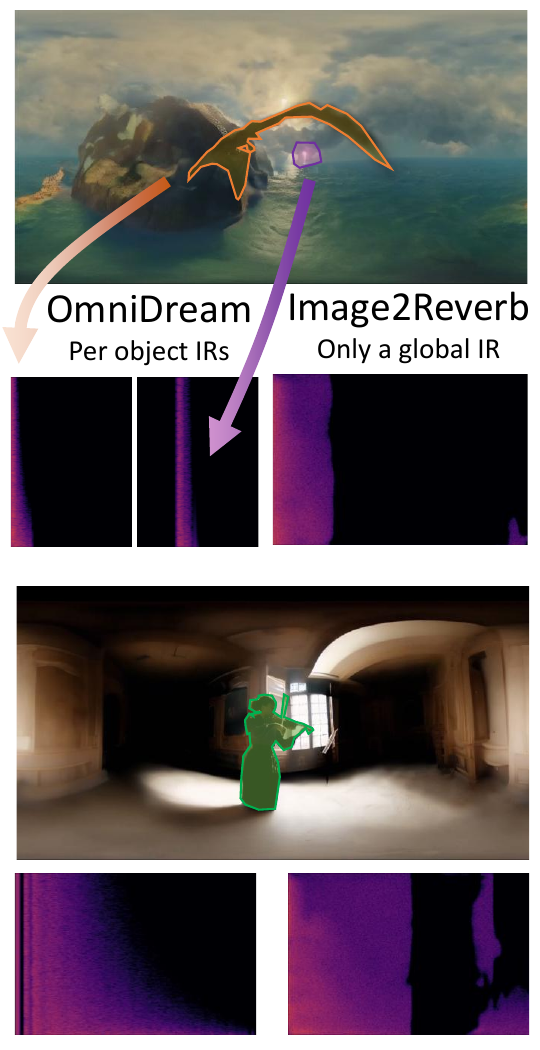}
    \vspace{-15pt}
    \captionof{figure}{Comparison of rendered impulse responses.}
    \label{fig:rir}
    \end{minipage}
    \vspace{-10pt}
\end{figure}

\begin{table}[t]
\centering
\scriptsize
\resizebox{0.8\linewidth}{!}{%
\begin{tabular}{l cc @{\hspace{8pt}} cc cc cc}
\toprule
 & \multicolumn{2}{c}{\multirow{2}{*}{AV Alignment}}
 & \multicolumn{6}{c}{Spatial Correctness} \\
\cmidrule(lr){4-9}
 & \multicolumn{2}{c}{}
 & \multicolumn{2}{c}{Overall}
 & \multicolumn{2}{c}{1\,fps}
 & \multicolumn{2}{c}{5\,fps} \\
\cmidrule(lr){2-3} \cmidrule(lr){4-5} \cmidrule(lr){6-7} \cmidrule(lr){8-9}
Method
  & IB$\uparrow$ & DS$\downarrow$
  & CC$\uparrow$ & AUC$\uparrow$
  & CC$\uparrow$ & AUC$\uparrow$
  & CC$\uparrow$ & AUC$\uparrow$ \\
\midrule
MMAudio                  & 0.25 & 0.89 & 0.11 & 0.55 & 0.11 & 0.52 & 0.11 & 0.52 \\
MMAudio (Spatial)        & 0.24 & 0.93 & 0.25 & 0.61 & 0.31 & 0.59 & 0.20 & 0.56 \\
See2Sound                & 0.06 & 1.08 & 0.05 & 0.35 & 0.09 & 0.37 & 0.10 & 0.38 \\
ViSAGe                   & 0.15 & 1.00 & 0.33 & 0.60 & 0.35 & 0.59 & 0.22 & 0.61 \\
OmniAudio                & 0.13 & 1.04 & 0.22 & 0.58 & 0.22 & 0.54 & 0.21 & 0.54 \\
 \midrule
\name{} (CubeComposer) & \textbf{0.27} & \textbf{0.87} & \textbf{0.50} & \textbf{0.72} & \textbf{0.49} & \textbf{0.70} & \textbf{0.46} & \textbf{0.68} \\
\textcolor{gray}{\name{} (Argus)} & \textcolor{gray}{0.26} & \textcolor{gray}{0.89} & \textcolor{gray}{0.48} & \textcolor{gray}{0.71} & \textcolor{gray}{0.46} & \textcolor{gray}{0.67} & \textcolor{gray}{0.43} & \textcolor{gray}{0.63} \\
\bottomrule
\end{tabular}
}
\vspace{-8pt}
\caption{
  Evaluation of video-to-spatial-audio generation on ImmerseSet.
}
\vspace{-20pt}
\label{tab:ours_results}
\end{table}

\begin{figure}[t]
    \centering
    \includegraphics[width=0.85\linewidth]{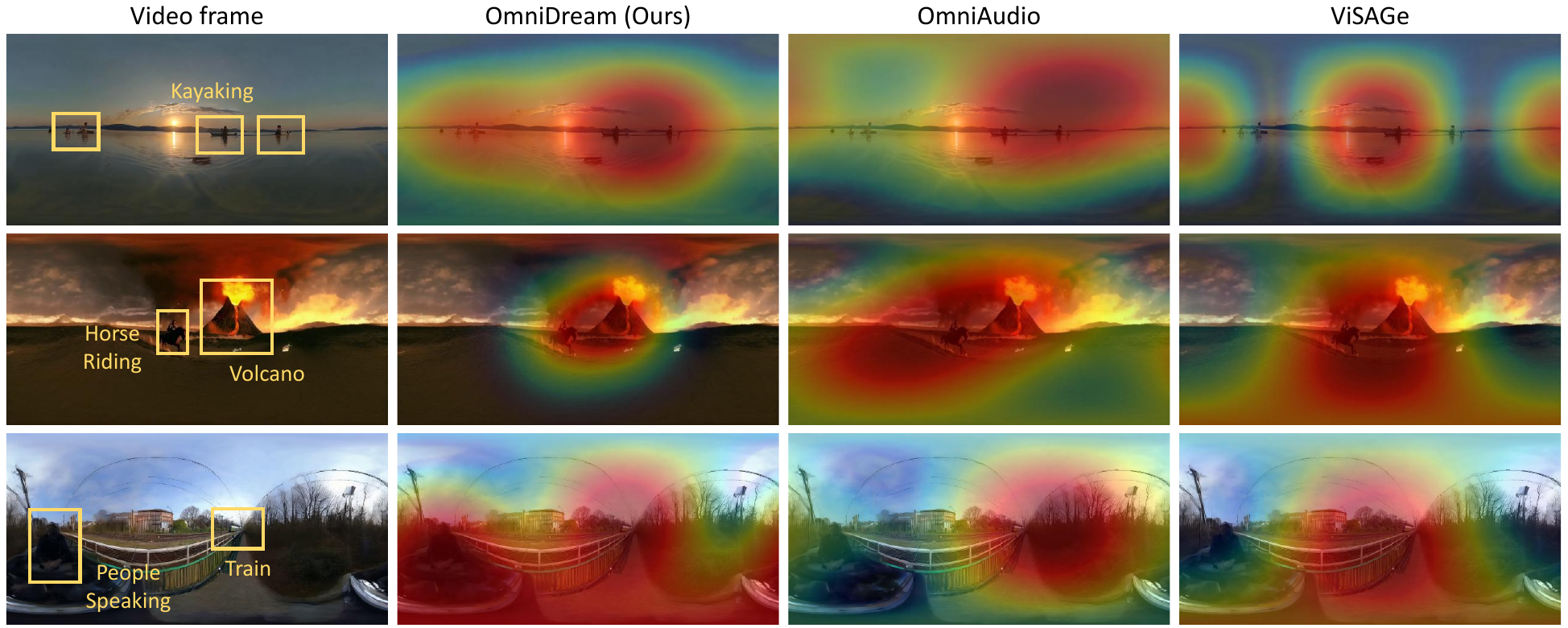}
    \vspace{-5pt}
    \caption{
        Spatial audio energy maps overlaid on equirectangular panoramic frames.
        Warmer colors indicate higher directional energy.
        \name{} concentrates energy on visible sounding objects, while baseline models produce incorrect spatial distribution.}
    \vspace{-10pt}
    \label{fig:energy_map}
\end{figure}

\begin{table}[t]
\centering
\scriptsize
\resizebox{0.85\linewidth}{!}{%
\setlength{\tabcolsep}{4pt}%
\begin{tabular}{l cc  cc cc cc  cc}
\toprule
 & \multicolumn{2}{c}{\multirow{2}{*}{AV Alignment}}
 & \multicolumn{6}{c}{Spatial Correctness}
 & \multicolumn{2}{c}{\multirow{2}{*}{Audio Similarity}} \\
\cmidrule(lr){4-9}
 & \multicolumn{2}{c}{}
 & \multicolumn{2}{c}{Overall}
 & \multicolumn{2}{c}{1\,fps}
 & \multicolumn{2}{c}{5\,fps}
 & \multicolumn{2}{c}{} \\
\cmidrule(lr){2-3} \cmidrule(lr){4-5} \cmidrule(lr){6-7} \cmidrule(lr){8-9} \cmidrule(lr){10-11}
Method
  & IB$\uparrow$ & DS$\downarrow$
  & CC$\uparrow$ & AUC$\uparrow$
  & CC$\uparrow$ & AUC$\uparrow$
  & CC$\uparrow$ & AUC$\uparrow$
  & KL$\downarrow$ & FD$\downarrow$ \\
\midrule
MMAudio
  & 0.23 & 0.76
  & 0.17 & 0.56
  & 0.23 & 0.57
  & 0.27 & 0.58
  & 2.09 & 8.59 \\
MMAudio (Spatial)
  & 0.22 & 0.79
  & 0.41 & 0.62
  & 0.44 & 0.64
  & 0.41 & 0.61
  & 2.07 & 8.60 \\
See2Sound
  & 0.05 & 1.31
  & 0.01 & 0.44
  & 0.02 & 0.42
  & 0.02 & 0.42
  & 2.96 & 18.53 \\
ViSAGe
  & 0.14 & 1.21
  & 0.38 & 0.61
  & 0.36 & 0.64
  & 0.33 & 0.63
  & 2.05 & 11.28 \\
OmniAudio
  & 0.15 & 1.22
  & 0.54 & 0.78
  & 0.50 & 0.76
  & 0.46 & 0.74
  & \textbf{1.43} & \textbf{5.50} \\
\name{}
  & \textbf{0.24} & \textbf{0.73}
  & \textbf{0.58} & \textbf{0.84}
  & \textbf{0.67} & \textbf{0.87}
  & \textbf{0.64} & \textbf{0.81}
  & 1.56 & 6.85 \\
\bottomrule
\end{tabular}
}
\vspace{-5pt}
\caption{
  Evaluation of video-to-spatial-audio generation on the YT360 dataset.
}
\vspace{-10pt}
\label{tab:yt360_results}
\end{table}

\header{Quantitative Results.}
Tab.~\ref{tab:ours_results} and~\ref{tab:yt360_results} report results on ImmerseSet and YT360 respectively.
On ImmerseSet, \name{} outperforms all baselines across every metric.
Spatializing MMAudio improves spatial correspondence, but coarse directional partitioning remains less effective than object-level decomposition and geometry-aware propagation. ViSAGe is limited by its fixed perspective view, while OmniAudio performs best among the baselines but still trails \name{}, showing that panoramic input alone is insufficient without explicit scene grounding.
Fig.~\ref{fig:example_visualization} visualizes one generated scene. As the user looks toward different perspective views, each sounding object produces a different spatial sound effect, and final audio is the sum of all object tracks.
On YT360, \name{} achieves the best performance across all AV alignment and spatial correctness metrics. OmniAudio obtains lower KL and FD, since its training corpus overlaps with YT360, although evaluation is conducted on a held-out split. Nevertheless, \name{} remains competitive in audio similarity while consistently outperforming OmniAudio in audiovisual alignment and spatial correctness.

\header{Spatial Audio Energy Maps.}
We visualize the spatial energy distribution in Fig.~\ref{fig:energy_map}.
Each cell in the energy map reflects the directional audio intensity decoded from the ambisonic signal.
Compared with the baseline models, \name{} concentrates audio energy around visually identifiable sounding objects: horse riding and a volcanic eruption in the second row, and a person paddling a kayak in the row.
This confirms that our object-centric decomposition and physics-grounded simulation correctly align generated audio with the panoramic video spatially.

\header{Acoustic Rendering.} \name{} simulates unique IR for each sounding object based on the recovered acoustic environments (Fig.~\ref{fig:rir}).
Concert halls (on the bottom) produce long reverberant tails with early reflections, while open outdoor scenes (top) yield short and dry responses since there are fewer acoustic reflections in the environments.
In the outdoor scene, the difference in arrival times between the dragon and the lightning stems from their relative distances to the camera.
Since the lightning is much farther away, its direct sound is significantly attenuated and arrives with a delay in the rendered IR.
In contrast, Image2Reverb~\citep{singh2021image2reverb} that directly conditions on the whole image can only generate one global IR with clear artifacts.
We additionally evaluate the accuracy of our acoustic rendering against measured real IRs in App.~\ref{sec:app_raf}.

\vspace{-5pt}
\subsection{Ablation and Analysis}
\label{sec:ablation}
\label{sec:robustness}
\label{sec:runtime}
\vspace{-5pt}

\header{Component Ablations.}
We ablate key design choices in Tab.~\ref{tab:ablation}. Ablation Variants include:

\begin{itemize}[leftmargin=10pt, itemindent=0pt, itemsep=2pt, topsep=-2pt, parsep=0pt]
    \item \textbf{Object-centric audio generation}: Instead of generating audio conditioned on per-object masked visual streams, we directly feed the full $360^\circ$ panoramic video to the audio generation model. We then apply spatial acoustic effects to the dominant sounding object in the scene.
    \item \textbf{Mask dilation}: We use  binary segmentation masks without morphological dilation, so each object tube captures only the object itself. This tests whether incorporating local environmental context improves audio generation.
    \item \textbf{Spatialization}: We replace the acoustic rendering with simple acoustic spatialization method including azimuth panning, with distance attenuation, and with propagation delay.
\end{itemize}

\begin{table}[t]
\centering
\scriptsize
\resizebox{0.85\linewidth}{!}{%
\begin{tabular}{l cc @{\hspace{8pt}} cc cc cc}
\toprule
 & \multicolumn{2}{c}{\multirow{2}{*}{AV Alignment}}
 & \multicolumn{6}{c}{Spatial Correctness} \\
\cmidrule(lr){4-9}
 & \multicolumn{2}{c}{}
 & \multicolumn{2}{c}{Overall}
 & \multicolumn{2}{c}{1\,fps}
 & \multicolumn{2}{c}{5\,fps} \\
\cmidrule(lr){2-3} \cmidrule(lr){4-5} \cmidrule(lr){6-7} \cmidrule(lr){8-9}
Variant
  & IB$\uparrow$ & DS$\downarrow$
  & CC$\uparrow$ & AUC$\uparrow$
  & CC$\uparrow$ & AUC$\uparrow$
  & CC$\uparrow$ & AUC$\uparrow$ \\
\midrule
w/o object tracking
  & 0.25 & 0.89 & 0.26 & 0.52 & 0.22 & 0.47 & 0.19 & 0.44   \\
w/o mask dilation
  & 0.23                 & 1.03 & 0.40 & 0.71 & 0.44 & 0.65 & 0.27 & 0.57  \\
\midrule
\multicolumn{9}{l}{ {Spatialization (replacing acoustic rendering):}} \\
azimuth panning
  & 0.21                 & 0.96 & 0.48 & 0.71 & 0.42 & 0.64 & 0.37 & 0.57   \\
\; + distance attenuation
  & 0.22                 & 0.93 & 0.49 & 0.70 & 0.45 & 0.65 & 0.41 & 0.63 \\
\; + propagation delay & 0.22                 & 0.94 & 0.49 & 0.71 & 0.45 & 0.65 & 0.41 & 0.62 \\
\midrule
\name{}
& \textbf{0.27} & \textbf{0.87} & \textbf{0.50} & \textbf{0.72} & \textbf{0.49} & \textbf{0.70} & \textbf{0.46} & \textbf{0.68} \\
\bottomrule
\end{tabular}
}
\vspace{-8pt}
\caption{
  Ablation studies on the ImmerseSet dataset.
}
\vspace{-18pt}
\label{tab:ablation}
\end{table}

Removing object-centric tracking causes the largest drop in spatial correctness, confirming that per-source decomposition is the primary driver of spatial accuracy; AV alignment is comparatively less affected since a global track can still loosely match dominant scene semantics.
Removing mask dilation produces the sharpest decline in AV alignment, indicating that local environmental context around each source is critical for temporal coupling between sound events and their visual triggers.
For the spatialization effect, adding distance attenuation or propagation delay to azimuth panning yields only a marginal gain.
Only the full renderer improves every metric, with its widest margins at the finer temporal resolution and in semantic alignment.
The benefit of physics-grounded rendering is therefore from scene-dependent acoustic simulation.

\header{Source Audio Dryness.}
Physics rendering assumes the per-object signals carry no or little acoustic effect.
Applying the same blind reverberation estimator to the generated sources, 90.16\% receive a predicted reverberation time below 0.15\,s, with a mean of 0.08\,s.
The generated object audio is therefore mostly dry, leaving little pre-existing acoustic effect for the simulated IR to conflict with.
The few sources scoring above the threshold are almost all repetitive sounds, such as crowds and ocean waves, which already contains complicated acoustic effect.
We therefore use a 0.15\,s reverberation-time threshold to screen sources before physics-based acoustic rendering.

\begin{figure}[!t]
\centering
\begin{minipage}[t]{0.38\linewidth}
\centering
\vspace{0pt}
\scriptsize
\captionsetup{hypcap=false}
\setlength{\tabcolsep}{2.2pt}
\begin{tabular}{l cc cc}
\toprule
Perturbation & IB$\uparrow$ & DS$\downarrow$ & CC$\uparrow$ & AUC$\uparrow$ \\
\midrule
Material        & 0.26 & 0.90 & 0.44 & 0.69 \\
Scene scale     & 0.27 & 0.88 & 0.45 & 0.67 \\
Source location & 0.26 & 0.89 & 0.37 & 0.62 \\
\midrule
\name{} & \textbf{0.27} & \textbf{0.87} & \textbf{0.5} & \textbf{0.72} \\
\bottomrule
\end{tabular}
\vspace{-4pt}
\captionof{table}{Robustness analysis.}
\label{tab:robustness}
\end{minipage}
\hspace{-0.01\linewidth}%
\begin{minipage}[t]{0.2\linewidth}
\centering
\vspace{0pt}
\scriptsize
\captionsetup{hypcap=false}
\setlength{\tabcolsep}{3pt}
\resizebox{\linewidth}{!}{%
\begin{tabular}{l r}
\toprule
\multicolumn{2}{l}{Video expansion (8s video)} \\
\;\; \textit{CubeComposer}               & 1680s \\
\;\; \textit{Argus}                      & 173s \\
Depth estimation            & 44s \\
Material segmentation       & 28s \\
Object tracking             & 88s \\
Audio generation            & 197s \\
Acoustic rendering          & 55s \\
\bottomrule
\end{tabular}
}
\vspace{-7pt}
\captionof{table}{Runtime.}
\label{tab:runtime}
\end{minipage}
\hspace{0.03\linewidth}%
\begin{minipage}[t]{0.38\linewidth}
\centering
\vspace{-2pt}
\captionsetup{hypcap=false}
\includegraphics[width=\linewidth]{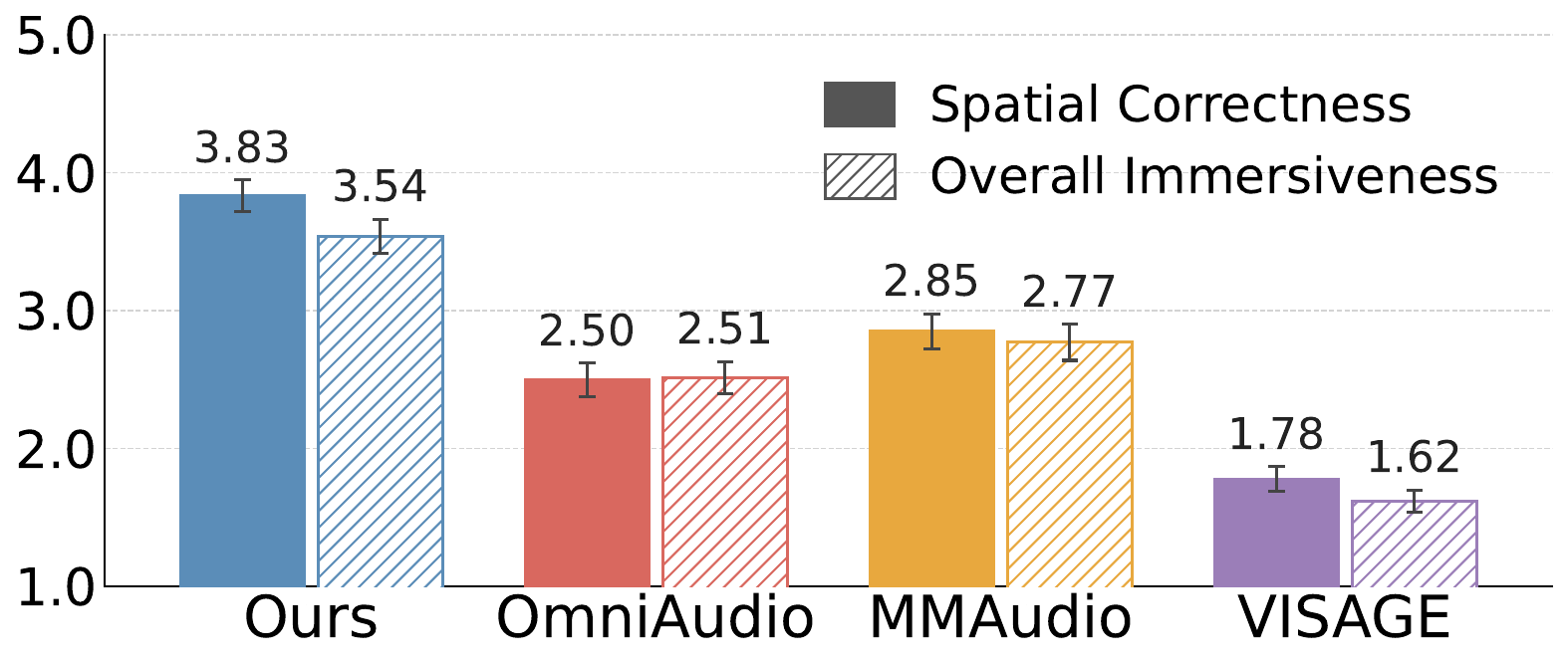}
\vspace{-20pt}
\captionof{figure}{Human evaluation.}
\label{fig:user_study}
\end{minipage}
\vspace{-20pt}
\end{figure}

\header{Effect of the Visual Frontend.} To evaluate the modularity of \name{}, we replace CubeComposer with Argus as the panoramic video frontend while keeping other components fixed.
As shown in Tab.~\ref{tab:ours_results}, CubeComposer performs consistently better across all end-to-end metrics.
CubeComposer also achieves the highest PSNR and SSIM and the lowest FVD in the standalone panoramic video comparison (Tab.~\ref{tab:video_quality}).
Together with the qualitative examples in App.~\ref{sec:app_qual}, these results support our choice of CubeComposer as the default visual frontend.

\header{Robustness to Scene Estimation Errors.}
Physics-grounded rendering relies on material, scale, and source-location estimates obtained from the input video. We therefore evaluate whether the complete system remains effective when these estimates are imperfect.
We independently perturb material reflection coefficients with Gaussian noise ($\sigma\!=\!0.12$), rescale the scene by $\pm20\%$, and perturb each source location with Gaussian noise ($\sigma\!=\!0.2$\,m per axis). Tab.~\ref{tab:robustness} reports the resulting end-to-end performance on ImmerseSet.
The system is robust to substantial errors in material and scale estimation.
Material perturbations retain a moderate level of performance degradation. Source-location noise has the largest effect.
This behavior is consistent with the evaluation: material and scale errors alter acoustic character, whereas localization errors directly move rendered energy away from the visual source direction.
Overall, the gains from physics-grounded rendering persist under plausible upstream errors rather than depending on perfect scene recovery.

\header{Runtime.}
Tab.~\ref{tab:runtime} reports per-stage costs on 8-second, 10\,fps clips with 3.4 sources on average, measured on an NVIDIA L40 GPU. Panoramic expansion takes about 1680\,s with CubeComposer, compared with 173\,s with Argus. This is a substantial runtime cost for the higher visual quality and stronger end-to-end results of CubeComposer (Tabs.~\ref{tab:video_quality} and~\ref{tab:ours_results}). Among the remaining stages, per-object audio generation takes 197\,s and can be parallelized across sources. Panoramic expansion is therefore the current speed bottleneck. Because \name{} has a replaceable visual frontend, faster and more capable panoramic video models can be adopted as they emerge without changing the audio generation or acoustic rendering stages.

\header{User Study.}
We conduct a human evaluation with \studynumber{} participants to assess perceptual quality along two dimensions: \textit{spatial correctness} and \textit{overall immersiveness}.
Participants listen to FOA audio decoded to binaural stereo via headphones while watching the corresponding $360^\circ$ video, and rate each output on a 1--5 MOS scale.
As shown in Fig.~\ref{fig:user_study}, \name{} achieves the highest scores on both dimensions, with a spatial correctness MOS of 3.83 compared to 2.85 for MMAudio and 2.50 for OmniAudio.
The gap is largest against ViSAGe (1.78), which struggles under the full $360^\circ$ setting as it was designed for perspective video.
On overall immersiveness, \name{} (3.54) again leads over MMAudio (2.77) and OmniAudio (2.51), confirming that physics-grounded spatialization translates to a perceptually stronger sense of presence.
Please refer to App.~\ref{sec:app_user_study} for more details on user study.

\vspace{-10pt}
\section{Discussion}
\label{sec:discussion}
\vspace{-10pt}

\header{Limitation and Future Work.}
Currently, \name{} does not support fully occluded sources, which lack the visual evidence our generation stage requires.
\name{} supports head rotation but not full free-viewpoint navigation. 
With further advances in 4D visual generation, future systems could support free viewer repositioning through novel-view synthesis across the full dynamic scene.
Looking ahead, we see these threads converging on a unified paradigm for immersive media generation, in which visual expansion, geometry, and spatial audio are produced jointly rather than stage by stage, absorbing existing mono soundtracks as conditioning, supporting streaming generation, and enabling end-to-end optimization once paired supervision exists for immersive media generation.
We regard our explicit decomposition as both a foundation and a baseline for that formulation.

\header{Conclusion.}
We presented \name{}, a framework that transforms a common silent video into an immersive $360^\circ$ audio-visual experience. 
Rather than treating immersive media generation as purely a visual or audio synthesis problem, \name{} jointly models panoramic scene reconstruction, object-centric sound generation, and physics-grounded acoustic propagation.
More broadly, our work highlights a new direction for generative media: transforming passive videos into explorable audio-visual worlds. 
We hope \name{} motivates future research at the intersection of multimodal generation, world model, 3D scene understanding, and interactive media.

\subsection*{AI use statement}
\vspace{-5pt}
ChatGPT was used to assist with language editing and manuscript preparation. All suggestions were reviewed by the authors, who take full responsibility for the paper's content. The proposed framework itself composes pretrained generative models, including CubeComposer for panoramic video generation, MMAudio for per-object audio generation, and the Gemini API for VLM-based sounding-object proposals and acoustic-property estimation. Their roles are described explicitly in the method and implementation details.

\subsection*{Ethics statement}
\vspace{-5pt}
Our user study involved only viewing immersive videos and listening to their generated audio and did not require IRB review under the applicable institutional guidelines. Participants were recruited on campus, received no compensation, were informed of the study procedure, and could withdraw at any time. We collected no personal or demographic information. All datasets were used under their applicable licenses. A potential societal risk is the misuse of generated immersive media to depict synthetic events as real; outputs should therefore be clearly disclosed as synthetic when presented outside research settings.

\subsection*{Reproducibility statement}
\vspace{-5pt}
We document the model configurations, prompts, preprocessing, hyperparameters, hardware, random seeds, runtime, and evaluation protocols in the paper and appendix. Upon acceptance, we will release the implementation, evaluation scripts, prompts and configurations, generated results, ImmerseSet resources, and interactive demo, together with instructions for obtaining the required pretrained models.

\bibliography{iclr2027_conference}
\bibliographystyle{iclr2027_conference}

\newpage
\appendix

\section*{Appendix}
\vspace{-5pt}
The appendix is organized as follows.
Sec.~\ref{sec:app_raf} studies the performance of physics-grounded acoustic rendering.
Sec.~\ref{sec:app_video_protocol} gives the protocol and results for the panoramic video-quality evaluation.
Sec.~\ref{sec:app_qual} presents additional qualitative results on the $360^\circ$ visual expansion stage and depth estimation.
Sec.~\ref{sec:app_foa} details FOA rendering, cross-fade mixing, and binaural playback.
Sec.~\ref{sec:app_user_study} describes the user study protocol.
Sec.~\ref{sec:app_impl} provides implementation details including model configurations, hyper-parameters, material acoustic property parameterization, and VLM prompts.
Sec.~\ref{sec:app_baselines} details how each baseline is configured.
Sec.~\ref{sec:app_metrics} describes the evaluation metrics in full detail.

\section{Evaluation of Physics-Grounded Acoustic Rendering}
\label{sec:app_raf}
\vspace{-5pt}

Our physics-grounded acoustic rendering essentially estimate the impulse response purely based on the visual scene. 
We therefore evaluate the accuracy of this component against measured impulse responses on Real Acoustic Fields (RAF)~\citep{chen2024real} dataset.

\header{Setup.} We use the furnished room of RAF, which provides densely measured real IRs together with a scanned room mesh. 
Instead of using dataset's ground truth geometry, we render a panoramic image from the provided mesh and then estimate everything the simulator needs with the same modules \name{} uses at inference: panoramic depth estimation for geometry, SAM2 segmentation with VLM querying for materials, and VLM metric-scale calibration.
Only the source and receiver positions come from the dataset, and the source is transformed by our estimated scale rather than the true one, so scale error propagates exactly as it would in deployment.

\header{Metrics.} All reported quantities are errors with respect to the measured IRs: C50 in dB, EDT in seconds, RT60 as a relative error, and a multi-resolution STFT error. Lower is better.

\begin{minipage}[t]{0.49\linewidth}
\centering
\vspace{0pt}
\small
\captionsetup{font=small,hypcap=false}
\setlength{\tabcolsep}{2.5pt}
\begin{tabular}{l cccc}
\toprule
Setting
  & \shortstack{C50}
  & \shortstack{EDT}
  & \shortstack{RT60}
  & \shortstack{STFT} \\
\midrule
Zero-shot  & 4.69 & 0.099 & 31.9 & 0.58 \\
One-shot   & 2.21  & 0.042 & 11.4 & 0.56 \\
INRAS  & 2.45  & 0.071 & 25.1 & 0.52 \\
\bottomrule
\end{tabular}
\vspace{-3pt}
\captionof{table}{Accuracy of IRs estimation. Zero-shot uses visual estimates only, one-shot uses one measured IR for calibration, and INRAS is trained on 300 samples.}
\label{tab:raf}
\end{minipage}
\hfill
\begin{minipage}[t]{0.49\linewidth}
\centering
\vspace{0pt}
\small
\captionsetup{font=small,hypcap=false}
\setlength{\tabcolsep}{2.5pt}
\begin{tabular}{l cccc}
\toprule
Perturbation & $\Delta$C50 & $\Delta$EDT & $\Delta$RT60  & $\Delta$STFT\\
\midrule
Material        & 0.74 & 0.0102 & 8.97\%  & 0.02\\
Scene scale     & 0.89 & 0.0163 & 13.46\% & 0.01 \\
Source location & 0.54 & 0.0089 & 6.06\% & 0.01 \\
\bottomrule
\end{tabular}
\vspace{-3pt}
\captionof{table}{Robustness of simulated IRs to errors in estimated materials, scene scale, and source location.}
\label{tab:rir_robustness}
\end{minipage}

\header{Results.} In the zero-shot setting used by \name{}, the simulated IR captures the overall time-frequency structure of the measured response, with an STFT error of 0.58, but does not fully recover its late-field energy, resulting in a 31.9\% RT60 relative error and a 4.69\,dB C50 error. 
As a diagnostic, we additionally use one measured reference IR to calibrate the room-level tail. 
This one-shot calibration reduces the C50 error to 2.21\,dB and the RT60 error to 11.4\%, matching or outperforming INRAS~\citep{su2022inras}, which is trained on 300 samples, while the STFT error changes only slightly from 0.58 to 0.56. This result indicates that the remaining error is concentrated primarily in late reverberation rather than the propagation structure, since a single measurement could not correct incorrect geometry.
One-shot calibration is not part of our pipeline and is used only for this diagnostic, as such a measurement is unavailable for arbitrary videos. 
We therefore claim that visually estimated geometry and materials provide a useful physical approximation rather than an exact acoustic reconstruction, and that this approximation is sufficient to improve immersive audio generation (Sec.~\ref{sec:ablation}).

\header{Robustness to estimation errors.} To test the robustness of the acoustic rendering pipeline, we perturb one visually estimated quantity at a time while holding the others fixed. Scene scale has the largest effect on the simulated response (13.46\% relative RT60 change) because rescaling modifies every propagation-path length. Material noise produces a smaller change (8.97\%), and source-location noise is mildest at the IR level (6.06\%). These results characterize how errors in the vision-derived scene representation propagate into the acoustic simulation; Sec.~\ref{sec:robustness} separately evaluates their effect on the complete immersive-generation system.


\section{Panoramic Video Generation Comparisons}
\label{sec:app_video_protocol}
\vspace{-5pt}

Since panoramic expansion supplies both the visual output and the geometry driving acoustic rendering, we quantify it directly against the three candidate backbones.

\header{Setup.} We use 30 YT360 clips, a corpus none of the three backbones trains on. Each backbone receives an identical perspective crop of the ground-truth panorama at $80^\circ$ FOV and $854\!\times\!480$ resolution, covering the leading 8\,seconds.
Because the methods generate at different native resolutions and lengths, no metric is computed on raw outputs: every output and the ground truth are resampled onto a common $1024\!\times\!512$ equirectangular grid and a common 27-frame window, 27 being the shortest native length so that no model is rolled out beyond its trained horizon.

\header{Metrics.} PSNR and SSIM are computed on the equirectangular frames, and FVD uses the StyleGAN-V I3D TorchScript extractor at the same resolution. The generated-region mask is the exact spherical footprint of the input frustum, computed by the pipeline's own conditioning routine rather than a rectangular latitude/longitude box, and covers 5.8\% of equirectangular pixels.

\begin{table}[h]
\centering
\small
\begin{tabular}{l ccc}
\toprule
Method & PSNR$\uparrow$ & SSIM$\uparrow$ & FVD$\downarrow$ \\
\midrule
CubeComposer~\citep{li2026cubecomposer} (ours) & \textbf{11.57} & \textbf{0.37} & \textbf{2417} \\
Argus~\citep{luo2025beyond}  & 11.13 & 0.34 & 2974  \\
PanoWan~\citep{xia2025panowan} & 8.19 & 0.22 & 5099 \\
\bottomrule
\end{tabular}
\vspace{-5pt}
\caption{Panoramic expansion quality on YT360 clips.}
\label{tab:video_quality}
\end{table}

\header{Results.} As shown in Tab.~\ref{tab:video_quality}, CubeComposer achieves the highest PSNR (11.57\,dB) and SSIM (0.37), and the lowest FVD (2417) among the three visual backbones. Compared with Argus, it improves PSNR by 0.44\,dB and SSIM by 0.03, while reducing FVD by 18.7\%. The improvements over PanoWan are larger: 3.38\,dB in PSNR, 0.15 in SSIM, and a 52.6\% reduction in FVD. These results are consistent with the qualitative comparisons in Sec.~\ref{sec:app_qual}. Together with its stronger end-to-end performance in Tab.~\ref{tab:ours_results}, they support our choice of CubeComposer as the default visual frontend for \name{}.



\vspace{-10pt}
\section{Additional Qualitative Results}
\label{sec:app_qual}
\vspace{-5pt}

\begin{figure}[h]
    \centering
    \includegraphics[width=\linewidth]{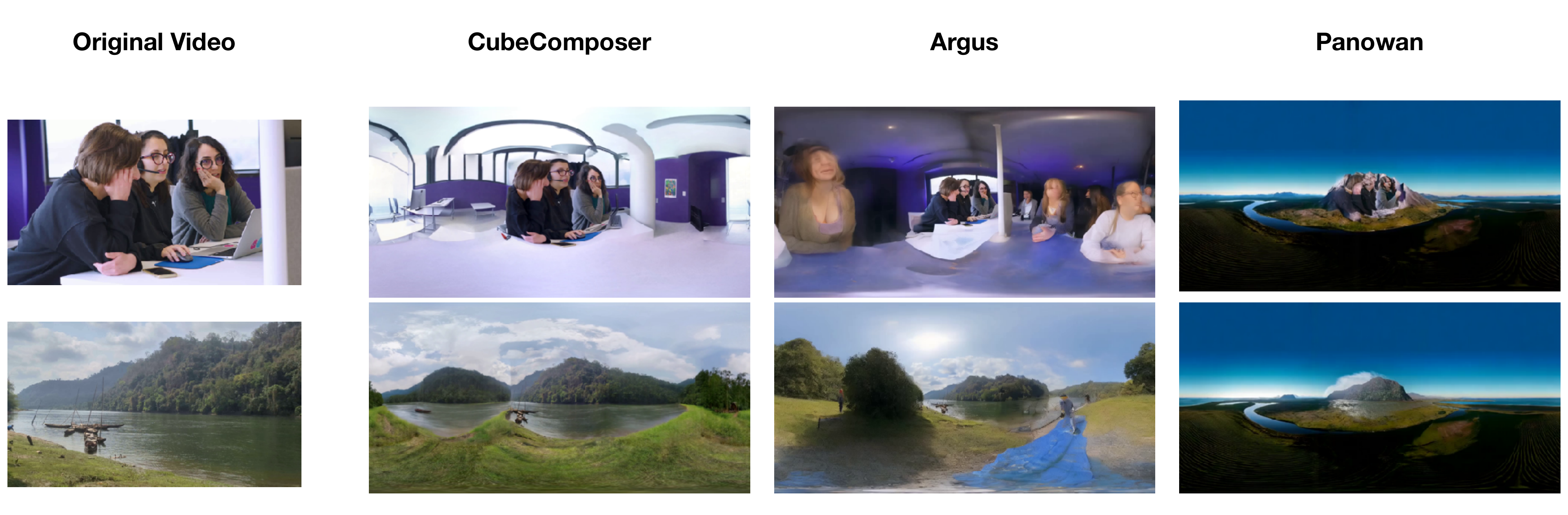}
    \caption{Comparison of perspective-to-$360^\circ$ video generation methods on two scenes. Each row shows the original perspective input (left) alongside the panoramic outputs of CubeComposer~\citep{li2026cubecomposer}, Argus~\citep{luo2025beyond},  and PanoWan~\citep{xia2025panowan}. CubeComposer generates the most coherent panoramic expansion, while Argus and PanoWan exhibit more visible inconsistencies in the generated regions.}
    \label{fig:app_visual_expansion}
\end{figure}

\header{$360^\circ$ Video Generation.} We compare three perspective-to-$360^\circ$ video generation methods including Argus~\citep{luo2025beyond}, CubeComposer~\citep{li2026cubecomposer}, and PanoWan~\citep{xia2025panowan}, on one indoor scene and one outdoor scene (Fig.~\ref{fig:app_visual_expansion}).
CubeComposer produces the most consistent panoramic expansions, better preserving the color palette and global scene structure of the original input. Argus and PanoWan exhibit more visible inconsistencies in the generated regions, including seam and warping artifacts that can propagate into depth estimation and acoustic simulation. These comparisons, together with the quantitative results in Tab.~\ref{tab:video_quality}, justify our choice of CubeComposer as the panoramic outpainting backbone.

\begin{figure}[h]
    \centering
    \includegraphics[width=\linewidth]{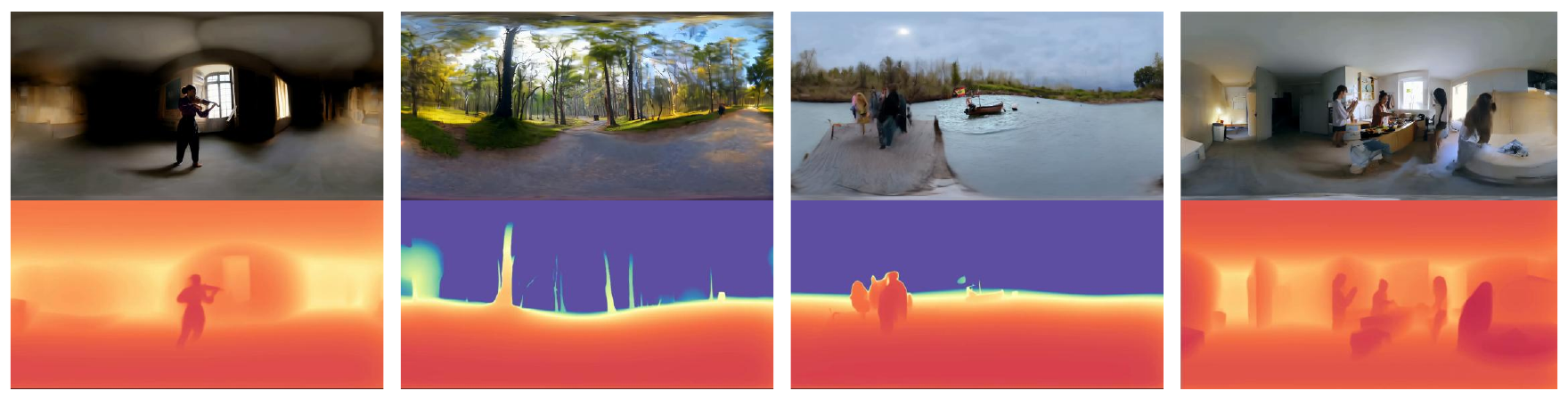}
    \vspace{-10pt}
    \caption{Example depth estimation for the generated panoramic videos.}
    \vspace{-10pt}
    \label{fig:depth_estimation}
\end{figure}

\header{Depth Estimation.} We evaluate panoramic depth estimation on the generated $360^\circ$ videos produced by our pipeline (Fig.~\ref{fig:depth_estimation}). Despite being synthesized through panoramic outpainting, the generated scenes preserve sufficient geometric consistency for downstream depth prediction. The estimated depth maps recover the dominant room layouts, open spaces, and large structural boundaries across both indoor and outdoor environments. These geometry estimates provide foundations for our acoustic simulation module, enabling physically grounded propagation and spatial audio rendering conditioned on the generated panoramic scenes.
\section{FOA Rendering, Cross-Fade Mixing, and Binaural Playback}
\label{sec:app_foa}
\vspace{-5pt}

Since both the scene geometry and source positions change over time, we simulate a separate four-channel FOA impulse response $\mathbf{h}_k^t \in \mathbb{R}^{4 \times L}$ for each source $k$ at every acoustic update $t$. In our implementation, updates occur once per second, and the IR is held fixed within each interval. Each dry source signal $s_k$ is divided according to these update intervals and rendered with the corresponding IR as described in Eq.~\ref{eq:foa_rendering}.

\header{Cross-fade mixing.} Directly concatenating rendered segments can introduce discontinuities at IR-update boundaries when the impulse response changes abruptly, for example when the source moves or the geometry changes. We therefore apply a raised-cosine cross-fade of length 300\,ms at each boundary. Let $C$ be the cross-fade length in samples, which is $C=13{,}230$ at 44.1\,kHz, and let $u\in\{0,\ldots,C-1\}$ index the overlap. The complementary windows are
\begin{equation}
\begin{aligned}
    w_{\mathrm{in}}[u] = \frac{1}{2}\!\left(1-\cos\frac{\pi u}{C-1}\right), \quad
    w_{\mathrm{out}}[u] = 1-w_{\mathrm{in}}[u].
\end{aligned}
\label{eq:crossfade_window}
\end{equation}
The two adjacent rendered segments are then combined by overlap-add:
\begin{equation}
\begin{aligned}
    \mathbf{a}_{k,\mathrm{xf}}[b_t+u]
    ={}w_{\mathrm{out}}[u]\,\mathbf{a}_k^t[b_t+u]+ w_{\mathrm{in}}[u]\,\mathbf{a}_k^{t+1}[b_t+u].
\end{aligned}
\label{eq:crossfade_mix}
\end{equation}
where $b_t$ is the start of the overlap between adjacent rendered segments. The windows sum to one at every overlap sample, smoothly transferring contribution between acoustic states. We apply the same scalar weights to all four FOA channels to preserve their directional relationships. After cross-fading, the rendered tracks are summed across sources to produce the final FOA mixture $\mathbf{A}$.

\header{Binaural playback.} During the user study, the viewer orientation $(\psi,\theta)$ is used to rotate the FOA field and decode it to $M$ virtual loudspeaker directions with an ambisonic decoding matrix $\mathbf{D}\in\mathbb{R}^{M\times4}$. Each decoded loudspeaker signal is then filtered with the corresponding left- and right-ear HRTFs:
\begin{align}
    \mathbf{Y}^t
    &= \mathbf{D}(\psi,\theta)\,\mathbf{A}^t,
    \label{eq:foa_decode}\\
    \mathbf{b}_{e}[n] = \sum_{m=1}^{M}\bigl(\mathbf{Y}_m^t &* h_m^e\bigr)[n],
    \quad e\in\{L,R\},
    \label{eq:binaural_decode}
\end{align}
where $h_m^L$ and $h_m^R$ are the HRTFs for virtual loudspeaker direction $m$. This decoding runs in real time so that the perceived sound field rotates consistently with the viewer.

\section{User Study Protocol}
\label{sec:app_user_study}
\vspace{-5pt}

\begin{figure}[h]
    \centering
    \includegraphics[width=0.85\linewidth]{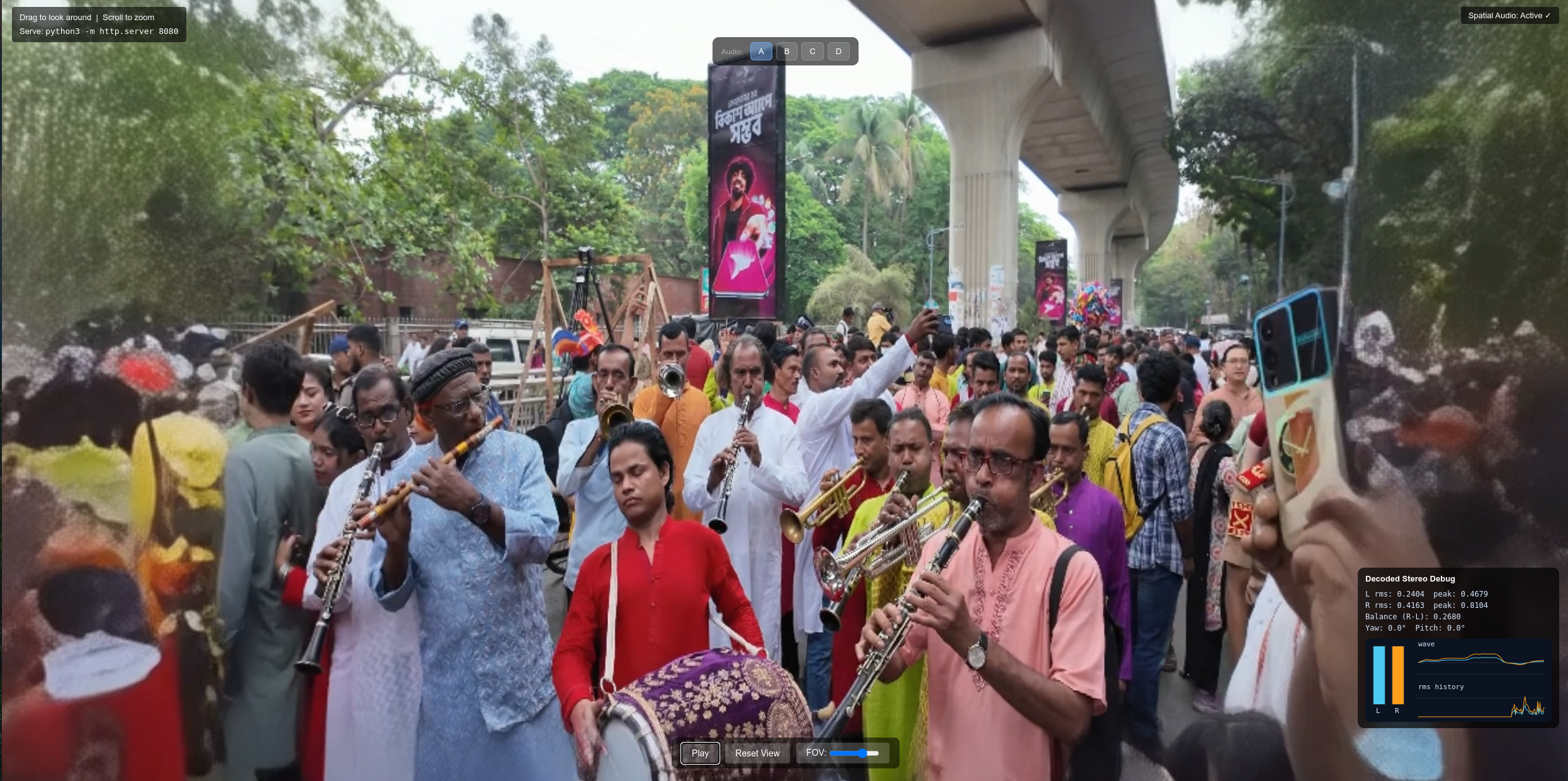}
    \caption{Interactive viewer for the user study. Participants drag to rotate the $360^\circ$ panoramic view; FOA audio is decoded to binaural stereo in real time as the viewpoint changes. The debug overlay (bottom right) shows live decoded stereo RMS/peak levels, and the current yaw/pitch orientation.}
    \label{fig:app_panel}
\end{figure}

\header{Interactive viewer.}
To deliver the immersive experience, we implement a browser-based interactive viewer in which the participant can freely drag to rotate the viewpoint within the $360^\circ$ panoramic video.
As the viewpoint changes, the FOA audio is decoded to binaural stereo following Sec.~\ref{sec:app_foa}, so the perceived sound directions update continuously with head rotation.
Fig.~\ref{fig:app_panel} shows a screenshot of the viewer interface, including a debug overlay that displays the live decoded stereo levels (L/R RMS and peak), channel balance, and current yaw/pitch orientation.

\header{Participants and rating procedure.}
We recruit \studynumber{} participants to watch and listen to our immersive media.
Each participant evaluates 6 clips per method (exclude See2Sound) in the ImmerseSet.
Method identity is hidden and randomized throughout the study.
Participants wear headphones and are instructed to freely explore the scene by dragging the view before and during rating.
After experiencing each clip, participants are asked to submit their evaluation on the google form with the following rating dimensions.

\header{Rating dimensions.}
Each clip is rated on a 1--5 MOS scale along two dimensions:
\begin{itemize}[leftmargin=10pt, itemindent=0pt, itemsep=2pt, topsep=0pt, parsep=0pt]
    \item \textbf{Spatial correctness}: does the audio appear to originate from the correct direction relative to the visual scene?
    \item \textbf{Overall immersiveness}: how strongly does the output create a sense of being present within the scene?
\end{itemize}

\section{Implementation Details}
\label{sec:app_impl}
\vspace{-5pt}

\header{Model configurations.}
\name{} is a training-free framework that composes the following off-the-shelf models.
Panoramic outpainting uses CubeComposer~\citep{li2026cubecomposer}.
We also alternatively use Argus~\citep{luo2025beyond} and PanoWan~\citep{xia2025panowan} for ablation studies.
Panoramic depth estimation uses~\citep{lin2025depth}, with metric scale estimated via Gemini~\citep{comanici2025gemini}.
Sounding object proposals and acoustic property assignment both use Gemini~\citep{comanici2025gemini}.
Object tracking and segmentation use SAM2~\citep{ravi2024sam}.
Per-object audio generation uses MMAudio~\citep{cheng2025mmaudio}.
Acoustic ray tracing is performed with AcoustiX~\citep{lan2024acoustic}.

\header{Hyperparameters.}
We use a mask dilation radius of $r\!=\!200$ pixels for object tube extraction.
The depth discontinuity threshold for mesh triangulation is $\tau\!=\!2$ meter.
Each per-frame acoustic simulation uses $10^6$ ray paths in AcoustiX.
And the sampling rate for the impulse response is set to be 44.1~kHz, which is consistent with the MMAudio~\citep{cheng2025mmaudio} output.
We sample $1$ keyframe per second when querying the VLM for sounding object proposals.
The impulse response simulation interval is set to 1 frame per second.
We simulate a new impulse response once per second and hold it constant within that interval.
The cross-fade window between adjacent frames is 300\,ms.
The runtime measurements in Tab.~\ref{tab:runtime} use an NVIDIA L40 GPU; the other inference experiments use a single NVIDIA B200 GPU.

\header{Acoustic material properties.}
\label{sec:app_impl_material}
For each segmented surface region, the VLM estimates three frequency-dependent acoustic properties across six octave-band centre frequencies: 500\,Hz, 1\,kHz, 2\,kHz, 4\,kHz, 8\,kHz, and 16\,kHz.
These are the reflection coefficient $\alpha_s^f \in [0,1]$, the absorption coefficient $\beta_s^f \in [0,1]$, and the scattering coefficient $\sigma_s^f \in [0,1]$, where $f$ indexes the frequency band and $s$ indexes the material class.
By design, $\alpha_s^f + \beta_s^f \approx 1$ at each frequency, as energy is either reflected or absorbed (transmission through thin surfaces is neglected).
The scattering coefficient controls the fraction of reflected energy that is diffusely scattered rather than specularly reflected, which governs late reverberation diffuseness.
This per-frequency parameterisation allows the ray tracer to capture frequency-dependent reverberation: hard dense surfaces such as glass ($\alpha \approx 0.97$ at 1\,kHz) sustain high reflection across all bands, while porous or soft materials such as carpet ($\alpha \approx 0.10$ at 1\,kHz) absorb more energy, especially at higher frequencies.
As noted in Sec.~\ref{sec:acoustic}, material properties are stable across frames for static surfaces, so the VLM is queried once per unique material label and the results are cached.

\header{VLM prompts.}
We use three separate VLM queries during the pipeline.

\textit{Metric depth calibration.}
In our pipeline, we treat the panoramic model's per-frame predictions as relative depth. To recover metric scale, we query the VLM with a panoramic frame and ask it to identify a single object with a known real-world size and estimate its average distance from the camera. The returned bounding box and depth estimate are used to compute a scale factor that maps the relative depth map to metric units.

\begin{promptbox}[Metric Depth Calibration Prompt]
\small
You are given a $360^\circ$ equirectangular panoramic image. Your task is to identify one object in the scene whose real-world size is well known (e.g.\ a person, a car, a door, a bicycle, a table), and estimate how far it is from the camera. \\[4pt]
Choose the object that you are most confident about in terms of both its identity and its distance. \\[4pt]
Return ONLY a JSON object with these fields: \\
\hspace*{8pt} \texttt{"object\_name"}: (string) name of the chosen object, e.g.\ \textit{"person"}, \textit{"car"}, \textit{"door"} \\
\hspace*{8pt} \texttt{"bbox"}: (list of 4 floats) bounding box of the object as [ymin, xmin, ymax, xmax] in 0--1000 coordinates \\
\hspace*{8pt} \texttt{"estimated\_depth\_m"}: (float) estimated average distance from the camera to the object centre, in metres \\
\hspace*{8pt} \texttt{"confidence"}: (float in 0--1) your confidence in the depth estimate \\[4pt]
Notes: \\
\hspace*{8pt} -- Prefer objects that are fully visible and unoccluded. \\
\hspace*{8pt} -- Use typical real-world sizes as anchors: a standing adult $\approx\!1.7$\,m tall; a standard car $\approx\!4.5$\,m long; a door $\approx 2.0$\,m tall. \\
\hspace*{8pt} -- If no object with a known size is visible, return \texttt{"confidence":0}.
\end{promptbox}

\textit{Sounding object proposal.}
We sample one keyframe per second from the panoramic video and pass all frames together to the VLM, which identifies persistent sounding objects across frames and reports each one's bounding box in the most representative frame.

\begin{promptbox}[Sounding Object Proposal Prompt]
\small
You are given $N$ frames sampled at 1 frame per second from a $360^\circ$ equirectangular panoramic video (full sphere unwrapped). \\[4pt]
Analyze all frames together to identify objects that plausibly produce sound throughout the video (e.g., people, musicians, animals, machinery, water, traffic). \\[4pt]
For each identified sounding object: \\
\hspace*{8pt} -- Track its presence across frames to confirm it is a genuine, persistent sound source. \\
\hspace*{8pt} -- Select the frame where it is most clearly visible as the representative frame. \\
\hspace*{8pt} -- Report its bounding box in that representative frame. \\[4pt]
Return JSON with an \texttt{"objects"} list, each entry having: \\
\hspace*{8pt} \texttt{"id"} (int), \\
\hspace*{8pt} \texttt{"name"} (str), \\
\hspace*{8pt} \texttt{"prompt"} (short sound description for audio generation, e.g.\ \textit{"crowd cheering"}), \\
\hspace*{8pt} \texttt{"representative\_frame"} (int, 0-indexed), \\
\hspace*{8pt} \texttt{"bbox"} ([ymin, xmin, ymax, xmax] in 0--1000 coordinates), \\
\hspace*{8pt} \texttt{"confidence"} (float in 0--1). \\[4pt]
Return at most 4 objects. Return an empty list if no sounding objects are found.
\end{promptbox}

\textit{Acoustic property assignment.}
Given segmented material regions from SAM2, the VLM is queried once per unique material label to estimate acoustic properties.

\begin{promptbox}[Acoustic Property Assignment Prompt]
\small
You are an acoustic engineering expert. This image shows a single surface region extracted from a $360^\circ$ equirectangular video frame. The background has been greyed out so you can focus on just this surface. \\[4pt]
Analyse the visible material and estimate its acoustic properties at six octave-band centre frequencies: 500\,Hz, 1\,kHz, 2\,kHz, 4\,kHz, 8\,kHz, and 16\,kHz. \\[4pt]
Return ONLY a JSON object with these fields: \\
\hspace*{8pt} \texttt{"material\_name"}: (string) specific material name, e.g.\ \textit{"polished marble"}, \textit{"rough concrete wall"} \\
\hspace*{8pt} \texttt{"description"}: (string) brief description of the surface and why it has these acoustic properties \\
\hspace*{8pt} \texttt{"frequencies"}: (list of int) [500, 1000, 2000, 4000, 8000, 16000] \\
\hspace*{8pt} \texttt{"reflection\_coefficients"}: (list of 6 floats) fraction of incident sound energy reflected at each frequency \\
\hspace*{8pt} \texttt{"absorption\_coefficients"}: (list of 6 floats) fraction of incident sound energy absorbed at each frequency \\
\hspace*{8pt} \texttt{"scattering\_coefficients"}: (list of 6 floats) fraction of reflected energy scattered diffusely at each frequency \\[4pt]
Notes: \\
\hspace*{8pt} -- \texttt{reflection + absorption} $\approx 1.0$ at each frequency (ignore transmission through thin surfaces). \\
\hspace*{8pt} -- Most soft or porous materials absorb more at higher frequencies. \\
\hspace*{8pt} -- Hard dense materials (glass, concrete, marble) maintain high reflection across all bands but may show slight absorption increase above 4\,kHz. \\
\hspace*{8pt} -- Typical references at 1\,kHz: glass $\approx 0.97$ reflection; carpet $\approx 0.10$ reflection; concrete $\approx 0.95$ reflection.
\end{promptbox}

\section{Baseline Configurations}
\label{sec:app_baselines}
\vspace{-5pt}

No existing method maps a silent perspective video to a $360^\circ$ video with FOA audio, so every baseline solves part of the task and must be given the input it was designed for.


\header{MMAudio}~\citep{cheng2025mmaudio}. A state-of-the-art video-to-audio model generating single-channel audio by latent diffusion. We run it on the generated $360^\circ$ video and replicate the single channel across four channels to form a pseudo-FOA signal. This carries no directional information, so its spatial scores reflect the absence of spatialization rather than a failure of the model, which was never designed to produce ambisonics.

\header{MMAudio (Spatial)}. Because channel replication is a weak spatial reference, we construct a stronger variant. We partition the panorama into four directional views, run MMAudio independently on each, and encode each resulting track into FOA at its view direction. This gives MMAudio genuine directional structure while still lacking object-level decomposition and scene geometry, isolating what those two components contribute.

\header{See2Sound}~\citep{dagli2025see}. An image-driven method that natively renders 5.1 surround audio from a single image. We condition it on the first frame of the generated $360^\circ$ video. Being image-based, it has no access to temporal dynamics, which is reflected in its synchronization scores.

\header{ViSAGe}~\citep{kim2025visage}. A vision-guided model predicting directional audio from perspective video. We apply it to the original perspective input, the setting it was designed for, rather than to the panorama. It therefore reasons only about the original field of view and cannot place sources in the surrounding $360^\circ$ scene.

\header{OmniAudio}~\citep{liu2025omniaudio}. A panoramic-video-to-FOA model. Since it requires $360^\circ$ input, we feed it the outpainted panoramic video $\mathbf{P}$ produced by our own visual expansion stage, so it receives the same visual context as \name{}'s audio stages and differs only in how spatial audio is produced.

\section{Evaluation Metrics}
\label{sec:app_metrics}
\vspace{-5pt}

\header{Spatial metrics.}
We adopt the correlation coefficient (CC) and AUC-Judd (AUC) metrics from ViSAGe~\citep{kim2025visage} to measure spatial correspondence between the generated audio and a reference spatial map.
For each frame, we decode the FOA signal into a directional energy map over the sphere and compare it against the reference map.
CC measures the linear correlation between the two maps, while AUC-Judd measures how well the energy map predicts high-reference locations. The reference map depends on the dataset. For \textbf{YT360}, which provides ground-truth FOA, we decode the reference energy map directly from the ambisonic audio following ViSAGe. For \textbf{ImmerseSet}, which has no ground-truth spatial audio, we construct a visually derived reference map from dense optical-flow magnitudes, aggregate it over each evaluation window, and resize and Gaussian-smooth it onto the equirectangular grid. We normalize both the predicted and reference maps before computing CC and AUC over the full clip and within 1\,s and 0.2\,s windows. The YT360 evaluation therefore measures agreement with reference acoustic energy, while the ImmerseSet evaluation provides a complementary measure of audio-visual spatial correspondence.

\header{AV synchronization metrics.}
We compute ImageBind score (IB) and DeSync (DS) using the open-source AV-Benchmark toolkit~\citep{cheng2024avbenchmark}, which is also used by MMAudio~\citep{cheng2025mmaudio}.
ImageBind score measures semantic alignment between audio and video in the shared ImageBind embedding space~\citep{girdhar2023imagebind}.
DeSync measures temporal misalignment as the absolute audio-video offset predicted by Synchformer~\citep{iashin2024synchformer}.

\header{Audio similarity metrics.}
KL divergence and Fréchet distance (FD) are computed using the same AV-Benchmark toolkit~\citep{cheng2024avbenchmark} between the distribution of generated audio features and ground-truth FOA recordings on the YT360 dataset.
These metrics are only reported on YT360, as ImmerseSet has no ground-truth audio reference.



\end{document}